# Grain Boundary Space Charge Engineering of Solid Oxide Electrolytes: Model Thin Film Study

T. Defferriere[1*], Y.B. Kim[2,3], C. Gilgenbach[1], J. M. LeBeau[1], W. Jung[2,3], H.L. Tuller[1*]

[1]Department of Material Science and Engineering, Massachusetts Institute of Technology, Cambridge, MA 02139, USA

[2]Department of Materials Science and Engineering, Seoul National University, Seoul 08826, Republic of Korea

[3]Research Institute of Advanced Materials, Seoul National University, Seoul 08826, Republic of Korea

*to whom correspondence should be addressed: tdefferr@mit.edu, tuller@mit.edu

ORCID: 0000-0002-3343-9226, 000-0001-8339-3222

1. Abstract

Grain boundaries (GB) profoundly influence the electrical properties of polycrystalline ionic solids. Yet, precise control of their transport characteristics has remained elusive, thereby limiting the performance of solid-state electrochemical devices. Here, we demonstrate unprecedented manipulation of space charge controlled ionic grain boundary resistance (up to 12 orders of magnitude) in metal oxide thin films. We exploit the orders of magnitude higher grain boundary diffusivities of substrate cation elements (i.e. Al from $Al_2O_3$ and Mg from MgO) relative to the bulk to modify the grain boundary chemistry, and thereby GB core charge, in a model oxygen ion conducting polycrystalline thin film solid electrolyte, Gd-doped $CeO_2$. This approach, confirmed jointly by TEM imaging and by extracting the respective GB and bulk diffusivities from measured SIMS profiles, enabled us to selectively control the chemistry of the GBs, while minimally modifying grain (bulk) chemistry or film microstructure, thereby ruling out potential effects of microstructure, strain or secondary phases. Broad tuning of GB space charge potentials is achieved by manipulating GB core charge density by over an order of magnitude, thereby providing a powerful tool for systematic studies of grain boundary phenomena across various functional materials. The implications of such control are far-reaching in achieving new functionality, improving efficiency and longevity of solid-state electrochemical devices.

## 2. Introduction

Grain boundaries (GB) are inherent features of polycrystalline materials, profoundly influencing the electrical behavior of functional materials. They contribute to reduced electronic mobility in semiconductors due to scattering, cause substantially higher GB resistances compared to the bulk due to the presence of carrier depletion zones adjacent to the interfaces, and serve as charge recombination sites in optoelectronic devices, limiting device sensitivity and efficiency[1]. In ion-conducting materials, GBs impede ionic transport and induce ohmic losses while also leading to chemical and mechanical instability under extreme loading conditions, degrading device performance over time[2,3]. In semiconducting ZnO-based varistors and $BaTiO_3$-based positive temperature thermistors, abrupt collapse or creation of space charge barriers at grain boundaries due to over-voltages or over-heating lead to desired switching behavior[4,5]. Given the ubiquity and the highly impactful effects of GB barriers to the flow of electronic and ionic charges, it is surprising how little success there has been in controlling their properties to achieve improved device performance and stability. In this study, we demonstrate the ability to *systematically* control the GB resistance of model oxygen ion conducting solid electrolyte thin films by up to 12 orders of magnitude. We achieve this by selectively engineering GB chemistry, without impacting the chemistry of the surrounding grains, thereby isolating the key role played by the magnitude and sign of the GB core charge.

The transport characteristics of GBs are highly sensitive to synthesis and processing conditions and depend on the nature and concentration of impurities, defect types, stoichiometric variations, and secondary phases localized at the interfaces. Even in high-purity systems, resistive GBs continue to be primarily induced by space charge effects that arise from charges trapped at GBs, creating electrostatic barriers that alter mobile charge carrier distributions in adjacent grains. The exact origin of the boundary core charge remains under debate, especially when dealing with ionic conductors[6–9]. Early models, inspired by semiconductors, point to carrier trapping at mid-gap defect states to explain GB core charge origins[1]. In ionic systems, these charges have been assigned to either intrinsic sources connected with free energy differences in the generation of, e.g., oxygen vacancies in oxygen ion conductors[7,9,10] or extrinsic effects due to impurity segregation to the GBs[8]. While common impurities like Al and Si have been correlated with increases in barrier height in ionic systems,[8] dopants used to enhance ionic conductivity in solid electrolytes, for instance, in $CeO_2$[11,12], $ZrO_2$[2,13,14], $(Ba, Sr)TiO_3$[15,16], $BaZrO_3$[17–20] and $BaCeO_3$[21], are known to reduce space charge barriers, thereby reducing GB blocking. Alkaline earth additives[22–26] and transition metals[11] used as sintering aids have also, at times, been reported to reduce GB space charge barrier heights in such metal oxide systems[5].

While our understanding of grain boundary structure and chemistry have advanced over the past decades, a direct link to space charge effects remains elusive. Unintentional impurities complicate the analysis, as they push the limits of detectability and are often limited to only a subset of interfaces. Nevertheless, studies indicate that residual impurities may critically impact space charge properties even in highly purified materials. Xu et al.[8,27], for example, showed that even in high purity (<40 ppm) undoped $CeO_2$ fibers and 0.2 at% Sm-doped $CeO_2$ bulks samples, grain boundaries exhibited unexpected levels of extrinsic impurities due to scavenging and accumulation of various cations (up to 2 at% Al for example). The impurity distribution within the same specimen correlated with a range of space charge potentials, some near zero, while some as high as 1 V, with similar observations being made for $SrTiO_3$[28,29]. These results indicate that grain boundary core charges in ionic solids may arise from trace impurities segregating to grain boundary regions.

Whether the origin of the GB core charge is intrinsic or extrinsic, methods for controlling the potential barriers are critical for achieving electrochemical devices with superior performance. Traditional grain boundary modification strategies rely on doping the host lattice and promoting dopant segregation during high-temperature treatments or sintering (typically > 1200ºC), but this makes it difficult to separate impacts of changing microstructure and grain chemistry from that of GB chemistry. Ideally, this could be circumvented by selective chemical modification of GBs following completion of microstructural development. Kwak et al.[30] previously demonstrated that GBs in polycrystalline $CeO_2$ films exhibit higher solubility and diffusion rates at moderate temperatures than bulk grains (600-800ºC), allowing for selective incorporation of elements into the GBs from a surface layer source via an in-diffusion thermal anneal step[31]. This allowed for controlled chemical modification of the GB interfaces without altering the bulk composition and minimally impacting the microstructure[31]. Systematic studies of how this influences space charge properties of grain boundary interfaces have nevertheless been limited both in extent and scope[33]. Litzelman et al.[32] previously observed that while Ni in-diffusion in $CeO_2$ thin films reduced space charge potentials, its impact was limited due to unexpected impurity up-diffusion from the MgO substrate. This observation is consistent with a report by Mills et al.[13] that suggested that the decreased grain boundary resistance in thin films of YSZ grown on MgO substrates at 700°C was due to a reduction in space charge potential, resulting from Mg diffusion into the YSZ film. Similar phenomena utilizing substrates with other compositions have not been investigated, leaving this approach's universality unexamined[33].

In this study, we demonstrate the ability to control grain boundary properties in a model oxygen ion conducting system, 3 at% Gd-doped $CeO_2$ ($Gd_{0.03}Ce_{0.97}O_{1.985}$ – GDC3). Our study relies on growing thin films of GDC3 on MgO and on $Al_2O_3$ substrates and subsequently exposing them to intermediate temperature anneals to induce selective grain boundary in-diffusion of the respective substrate elements. Fig. 1a illustrates the impact of excess positive charge, and the corresponding impact on the spatial dependence of the space charge potential $\phi_{sc}(x)$, the mobile concentration of doubly charged oxygen vacancies $[v_O^{\cdot\cdot}]$, the minority electron density $n$, and the bulk acceptor dopant density $[A'_{Ce}]$, assumed to be immobile at the lower annealing temperatures. Fig. 1b (upper section) illustrates negatively charged species on the left and positively charges species on the right selectively diffusing up a grain boundary, leading to respectively a net decrease (increase) in the net positive charge at the grain boundary. This leads in turn (see lower section Fig. 1b) to a decrease (increase) in $\phi_{sc}(x)$, a decrease (increase) in depletion of $[v_O^{\cdot\cdot}]$, and a decrease (increase) in the accumulation of minority electrons, $n$. With this approach we can achieve up to ~ 12 orders of magnitude change in GDC3's room temperature extrapolated resistance when compared against nominally epitaxial standards (GB resistance free). STEM and Secondary Ions Mass spectroscopy are used to verify the selective nature of our grain boundary chemical modification, while XRD and SEM studies confirm that this is achieved via minimal difference in bulk chemistry, strain, or microstructure. Our approach offers clear evidence for chemical control of grain boundary space charge properties and establishes guidelines that can be applied to solid electrolytes, thus offering a pathway to enhanced performance for polycrystalline ionic conductors.

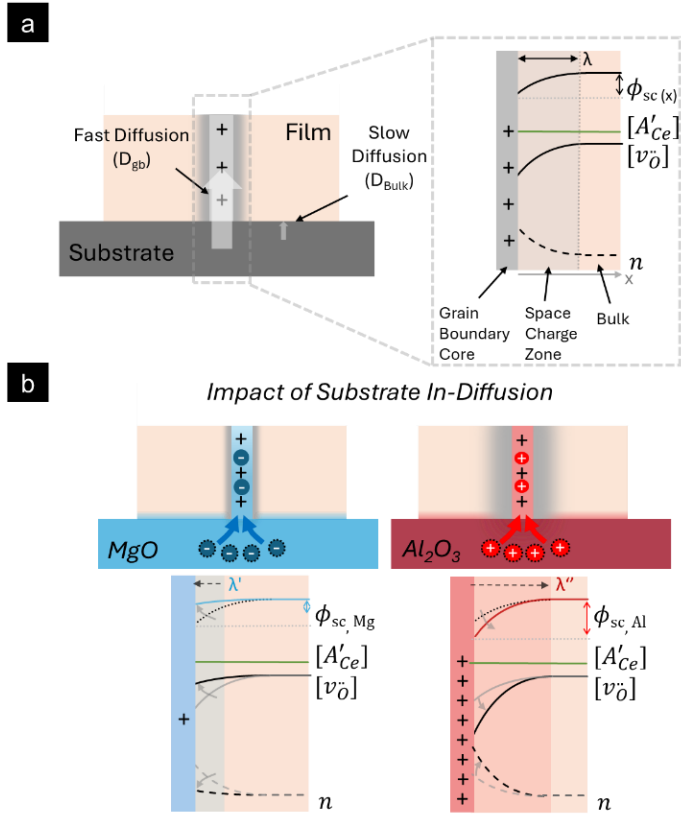

*Figure 1a). Illustration of polycrystalline film on substrate with a columnar GB, initially positively charged. GB enables selective up-diffusion of substrate cations into film. Inset is schematic of space charge potential $\phi_{SC(x)}$ profile generated by the positively charged GB core and corresponding majority oxygen vacancy ($V_o^{\cdot\cdot}$) depletion and minority electron n accumulation, following Mott Schottky model whereby dopant concentration $[A'_{Ce}]$ is assumed to be flat throughout space charge region. b) Illustration of impact of selective grain boundary in-diffusion on GB space charge potentials. Magnesium ions ($Mg^{2+}$), in blue, and aluminum ions ($Al^{3+}$), in red, respectively decrease and increase the density of positive core charge, modifying grain boundary space charge potentials, space charge width λ, and degree of ionic carrier $V_o^{\cdot\cdot}$ depletion (see text for details).*

## 3. Results

Polycrystalline GDC3 thin films (~200 nm thick) were prepared by pulsed laser deposition on (0001) $Al_2O_3$ and on (001) MgO single-crystal substrates, then exposed to 6 h thermal annealing conditions at temperatures ranging from 650 to 1250°C, and subsequently studied electrically by electrochemical impedance spectroscopy. The selection of the substrates was based on the ionic sizes and charges of the cations that make up the substrates relative to the size and charge of the GDC host cation $Ce^{4+}$ with the aim of modulating the net GB core charge (see methods section and equation S4 in SI). The details regarding sample fabrication and electrode geometry are provided in the experimental section and supplementary **section 2**, while the microstructural characterization can be found in supplementary **sections 3 & 4**. As confirmed by SEM and TEM (**Fig.** S1 & S2), the as-prepared samples are polycrystalline with columnar grains perpendicular to the substrate, with an initial (111)/(100) mixed grain orientation distribution. This is expected to facilitate the selective diffusion of substrate elements up the grain boundary channels. As shown in **Fig**. S3 and S4, XRD and surface SEM analysis show no differences in structural evolution with annealing temperatures for films grown on the different substrates and no evidence of changes in the presence of secondary phases or strain state which supports the idea that our electrical trends result solely from chemical modifications of the space charge properties.

*Electrical Conductivity Behavior:* In **Fig.** 2(a,b), the inverse resistance (1/R) x temperature (T) extracted from impedance measurements (characteristic Nyquist plots measured at 350°C shown in **Fig**. S6 in the SI) of all films grown on MgO and $Al_2O_3$ are plotted as a function of inverse temperature $(k_BT)^{-1}$ on an Arrhenius plot to enable extraction of corresponding activation energies. The conductance behavior of the films can be described, for the most part, as exhibiting single slopes. Deviations, however, appear at the lower temperatures due to expected current leakage along the film surface or the film/substrate interface, given the very high recorded resistance values (> GΩ) at the reduced temperatures. Focusing on the higher temperature behavior (> 150-200ºC), where clear trends are observable, and single slope activation energies are present, we find that the nominally epitaxial samples show the highest conductance's and lowest activation energies (~0.67 eV), consistent with other literature reports both on thin film and bulk ceramics, with the activation energy ascribed to bulk conductivity resulting from lattice oxygen vacancy migration[12,33]. The polycrystalline films exhibit larger activation energies than those of the epitaxial films (0.67eV), consistent with grain boundary resistance dominance and space charge related phenomena.

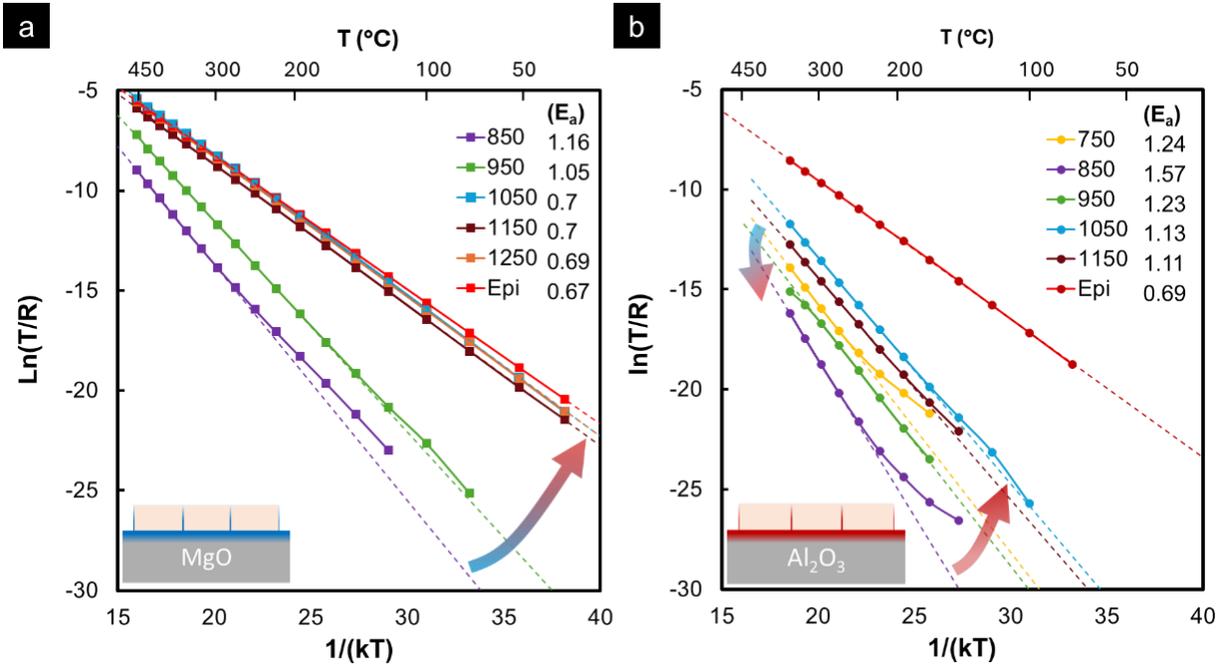

*Figure 2 (a,b) Natural logarithm plot of the Arrhenius dependence of conductance G as (1/R) x T vs. 1/$k_B$T derived from the impedance plots for the films grown on MgO and $Al_2O_3$ substrates, respectively. Activation energies ($E_a$ - in eV) obtained from slops of curves are displayed in legend.*

When examining the temperature dependence of film conductance grown on MgO, as displayed in **Fig.** 2(a), one observes that films annealed at the lowest temperature (850ºC) exhibit the highest activation energy (1.16 eV) and lowest conductance over the entire temperature range. The values reported are consistent with other literature reports both on thin films and bulk ceramics, ascribed to blocked oxygen ion transport at the GBs[12,33,34]. With increasing annealing temperature, the conductance (1/R) increases, while the activation energy decreases, with the total activation energy dropping from 1.16 eV to 0.67 eV, the latter aligning with the values characteristic of the nominally epitaxial sample. Likewise, the conductance increases toward that of the nominally epitaxial films. Above 1000ºC, the conductance and activation energies reach values close to those of the nominally epitaxial films, indicating a saturation of the observed phenomena.

The trend is less straightforward in the case of films grown on Al$_2$O$_3$. Starting at the lowest temperature anneal (750ºC), the films exhibit a high resistance and activation energy (~1.24 eV), slightly higher than the values obtained for the film grown on MgO. With increasing annealing temperature up to 850ºC, however, we observe the resistance and activation energy increasing substantially, with E$_a$ reaching a value of ~1.57 eV. Above 850ºC, the resistance and its activation energy, however, begin to decrease, reaching a saturation point > 1050ºC. We note that the conductance and activation energies, however, do not coincide with the values of the nominally epitaxial film, as observed for the films grown on MgO. Interestingly, the activation energy reaches a saturation value of (~1.15 eV), a value often reported for GDC oxides in thin film and bulk form following high-temperature sintering.

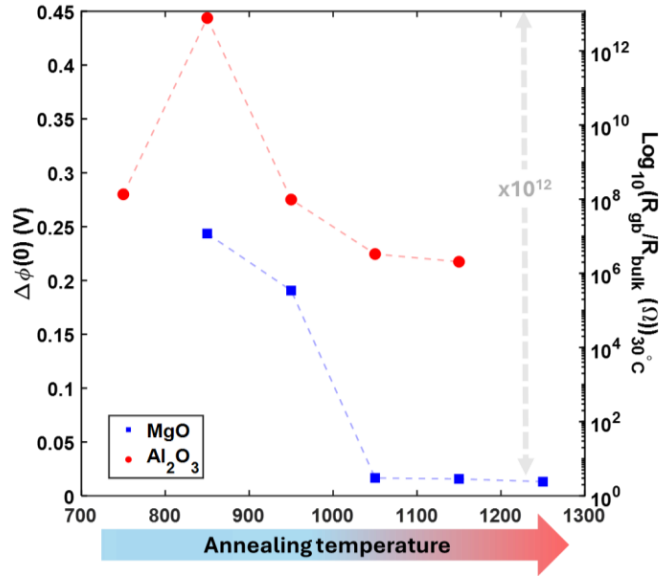

*Figure 3 Calculated space charge potential on left axis and extrapolated resistance ratios at 30°C on right axis. Results are obtained by comparing impedance of polycrystalline films with epitaxial ones (see section 7 in SI for procedural details) and using equation S2 derived from Mott Schottky approximation (section 1 in SI). All data in figure plotted vs annealing temperature for films grown on MgO vs Al$_2$O$_3$ - see text for details.*

As shown in **Fig.** 3, if we extrapolate the film conductivities down to room temperature based on their high-temperature slopes, the changes in film conductance with annealing temperature reach factors of $\geq 10^5$ in both cases. However, at its peak, around 850ºC, the conductance's of both films also differ by a factor of $10^5$. Using equation (S2), we obtain estimated values for the changes in space charge potential as displayed in **Fig.** 3 (procedure outlined section 7 of SI and plots showing fits and table with space charge potential values provided in **Fig.** S8). One observes that the space charge potential follows the same trends as the total activation energy and resistivity with annealing temperature. For the MgO case, the potential barrier begins at 0.25 V and decreases to near ~ zero (value 0.013V) with annealing, while for films grown on $Al_2O_3$, it begins at 0.27 V, reaches values as high as 0.45 V and finally decreases back down to 0.23 V, similar to the starting point for films on MgO. Based on these space charge values and applying equation (S4) that relates barrier height $\Delta\phi(0)$ to GB core charge Q, we can estimate that the average "net" core charge density ($Q_{core}$) at the grain boundaries changes by a factor ~ 6 from 0.047 to 0.276 C/m² over the entire range of conditions examined.

*Analysis of Depth Profiles by TOF-SIMS*: To elucidate the role that chemical modification has on the electrical properties of the grain boundaries; secondary ion mass spectroscopy (SIMS) was performed on the samples to measure the in-diffusion profiles of the substrate cations into the grain boundaries of the supported GDC films. **Fig.** 4(a) displays characteristic SIMS spectra for the dopants and host elements for samples as deposited and subsequently annealed at 750, and 850°C for 6 h, respectively. In SI section 8, we provide SIMS plots for samples annealed from 650 to 1250°C in the case of MgO and 750 to 1250°C for the case of $Al_2O_3$. In general, the depth-independent signals of Ce+ and Gd+ originate from the deposited GDC film, and $Mg^+$ and $Al^+$ each arise from the underlying MgO and $Al_2O_3$ substrates. In contrast to the as-deposited samples, where both Mg and Al signals drop sharply at the film-substrate interface, all samples that were annealed show significantly increased Mg or Al signal intensity to much deeper penetration depths. As shown in **Fig.** 4(a), the concentration profiles for Mg and Al appear to saturate by the time the samples' annealing temperature reaches 850°C, where the number of in-diffused species presumably reaches the solubility limit of the grain boundaries of the deposited GDC films. Above 850°C, as displayed in **Fig**. S9 in the SI, the saturation concentration in the bulk of the films decreased slightly with increasing annealing temperatures, whereas samples deposited on MgO showed an increase in Mg concentration near the surface. Such variation can be rationalized by noting that the decrease in grain boundary concentration is likely due to grain growth that eventually lowered the apparent grain boundary solubility per surface area, while the grooving/dewetting, especially at higher temperatures, enables additional surface diffusion to occur along the film surface, consistent with previous reports[13]. Discrepancies between Al and Mg may relate to a fundamental difference in surface diffusivity, which is beyond the scope of this article.

The dominant in-diffusion pathways were identified by fitting the diffusion profiles using both bulk and grain boundary diffusion expressions, detailed in SI section 8, with the results plotted as a function of temperature shown in **Fig**. S16. We note that the activation energies for diffusion along the grain boundaries are $\frac{1}{2}$ and $\frac{1}{5}$ of the bulk values ($Al_2O_3$ 1.3 eV vs 2.6 eV and MgO 0.84 eV vs 4.1 eV). We calculate the differences in diffusivities for the bulk vs grain boundaries at a common temperature, e.g. 850°C, through extrapolation. We find that $D_{GB,Al/Mg,850} \sim 10^{-16} cm^2/s$, while $D_{Bulk,Al/Mg,850} \sim 10^{-21} cm^2/s$, equating to ~5 orders of magnitude enhancement in Al and Mg cation diffusivities between grain boundary vs bulk, consistent with previous reports on markedly enhanced grain boundary diffusivity of foreign cation elements[30].

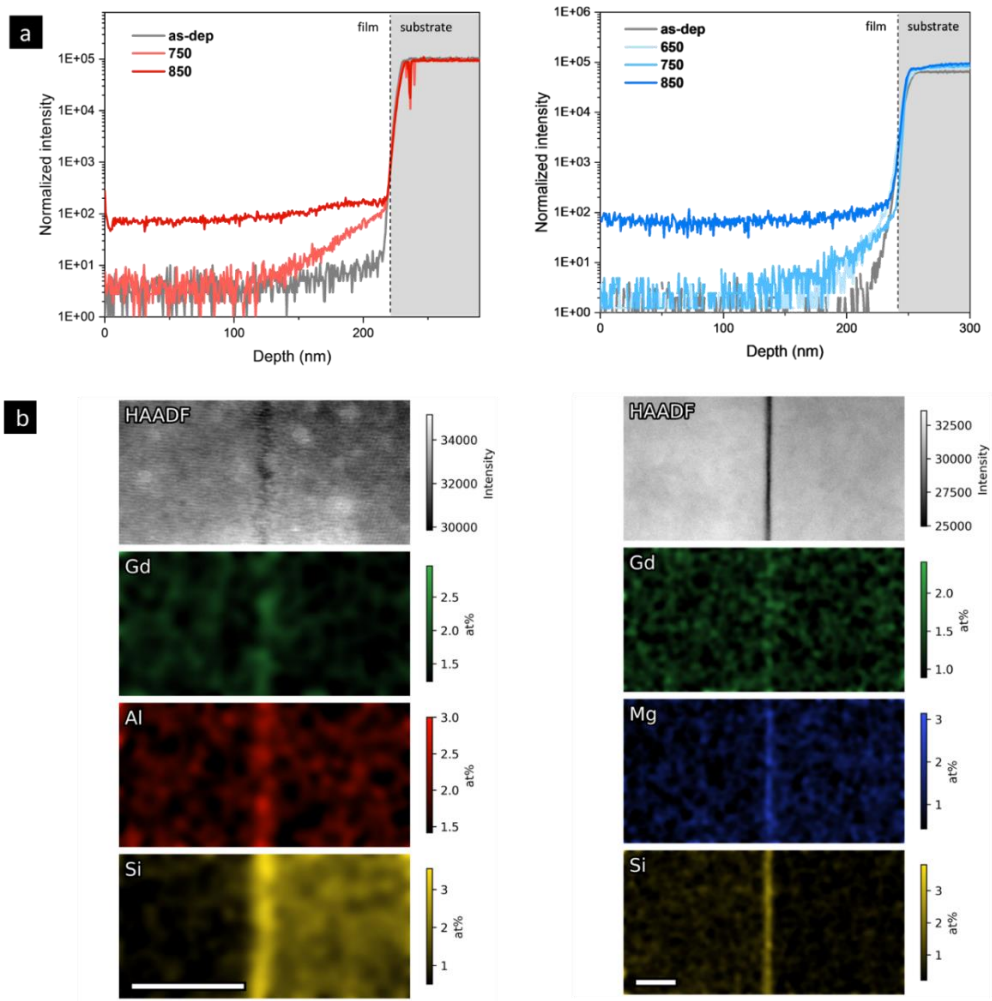

*Figure 4* (a) ToF-SIMS depth profiles of $Al^+$ in GDC films deposited on $Al_2O_3$ (left image) and $Mg^+$ in GDC film deposited on MgO (right image) obtained following film deposition and subsequent annealing treatments at 750 & 850°C for 6 hours for $Al_2O_3$ and 650, 750 and 850°C for MgO. The vertical dotted line signifies the interface between the GDC film and substrate, corresponding to the point where the signals of $Ce^+$ and $Al^+$/$Mg^+$ overlap. (b) STEM-EDS spectra images collected from a representative grain boundary in GDC grown on $Al_2O_3$ (left image) and on MgO (right image) at 1050°C. The top image displays HAADF intensity, while the bottom images show the semiquantitative concentrations of Gd, Mg, Al and Si. The Mg boundary shows strong enrichment of Mg, while the Al grain boundary shows evidence of Al enrichment with both showing evidence of Gd enrichment. In all cases, the enrichment is localized to within ~1 nm of the grain boundary core. Scale bars 10 nm.

*Analysis of Grain Boundary Segregation by STEM-EDS*: Scanning transmission electron microscopy (STEM) energy dispersive x-ray spectroscopy (EDS) corroborates the SIMS measurements, showing in-diffusion from the substrates into the GDC films as in **Fig**. 4(b) for samples annealed on alumina and magnesia at 1050°C. Additional STEM EDS line plots are present in **Fig**. S19, S20, S21 for samples on alumina annealed at 850°C and 1050 °C, and in **Fig.** S22 and S23 for samples on MgO annealed at 1050°C and 1250 °C). In both cases, the substrate cation is found to be enriched at the grain boundary and more pronounced at higher annealing temperatures.

In both cases, there is additionally evidence of Gd segregation to grain boundaries—this may also lower the measured space charge potentials (see discussion). Further, Si contamination was also found at the grain boundary (**Fig**. S21, S22, and S23) for samples annealed at higher temperatures (>1000°C), though no significant presence was found when analyzed by SIMS (see SI **Fig.** S17 and S18). The observation of Si contamination seemed to correlate with higher temperature anneal, as little evidence of Si was present in GDC3 samples grown on $Al_2O_3$ annealed at 850°C as shown in **Fig**. S19. In general, no other significant traces of potential contaminants (Ca, Na, H, Sr) could be found in either the SIMS or TEM results (see **Fig.** S17 and S18). Certain boundaries show little or no segregation and appear to be low-angle grain boundaries (see **Fig**. S20). This highlights the coupling between grain boundary structure and the resultant electrochemical properties.

## 4. Discussion

Our results demonstrate the ability to systematically and markedly vary the GB resistance in a model polycrystalline oxygen ion conducting solid electrolyte thin film ($Gd_{0.03}Ce_{0.97}O_{1.985}$). This leads to changes in room temperature extrapolated values by up to 12 orders of magnitude by controlling the net core charge density (positive – in this case) of the grain boundaries - as illustrated in **Fig.** 3. This was achieved by growing GDC3 thin films on both $Al_2O_3$ and MgO substrates and annealing them over a broad temperature range (750-1250ºC). We could rule out potential contributions from structural variations, as the same evolution was observed on both substrates regarding grain size, morphology, and grain orientation. Strain and secondary phases were furthermore ruled out by XRD and TEM studies. Since we were able to confirm from SIMS profile fitting that both $Al^{3+}$ and $Mg^{2+}$ ions diffuse ~$10^5$ times faster along the GBs than in the GDC grains, we were able to selectively and controllably modulate the grain boundary chemistry, and therefore the GB core charge, while maintaining the "bulk" grain chemistry constant.

The reader is reminded that GDC3 polycrystalline films, as grown, already exhibited a built-in space charge potential of ~ 0.25 V, consistent with a positive core charge (see detail in SI section 1) as previously reported in the literature[2,12,33,34]. Our electrical, SIMS and TEM results are all consistent with the assumption that in-diffusion of $Al^{3+}$ results in a net increase in positive charge, whereas in-diffusion of $Mg^{2+}$ results in a net decrease in positive charge at the GB core resulting in an overall change in net positive charge by a factor of ~ 6 (0.047 to 0.276 C/m²). These results align with the expectation that the much smaller $Al^{3+}$ ion (0.535 Å – 6 fold coordination) enters interstitially into sites within the grain boundary regions, leading to a net positive defect $Al_i^{\cdot\cdot\cdot}$, while the $Mg^{2+}$ ion (0.89 Å – 8-fold coordination) with a net negative charge $Mg_{Ce}''$, and with comparable size to the $Ce^{4+}$ ion (0.97 Å – 8 – fold coordination), enters substitutionally.[35]

While this explanation is consistent with the observed trends upon heating up to 850ºC, the reversal in trends for the Al in-diffused films above that temperature requires further explanation. It is instructive to examine the STEM-EDS spectrums that show that the net negatively charged bulk dopant $Gd'_{Ce}$ begins to significantly segregate to the grain boundaries above 850ºC (see Fig. S19-S23 in SI). This is expected to result in: *i)* the addition of negative charges within the grain boundary core and thereby *ii)* a decrease in the space charge width. In the case of the MgO substrate, the redistribution of both $Mg''_{Ce}$ and $Gd'_{Ce}$ to the grain boundary core together contribute to the decrease in space charge potential, while in the case of Al$_2$O$_3$, $Gd'_{Ce}$ and $Al_i^{\cdot\cdot\cdot}$ compete. This explains why Mg in-diffused films exhibit a monotonic decrease in GB resistance, while Al in-diffused films exhibit an initial decrease in conductivity at reduced temperatures and then an increase at higher temperature anneals. At the lower temperatures, only Al diffusion along the grain boundaries actively contributes to increased positive GB core charge. Then at higher temperatures during which the bulk dopant $Gd'_{Ce}$ becomes sufficiently mobile even within the grain, including within the space charge region, to enable it to diffuse the short distance to the GB core, at which point its negative charge begins to compensate for the extra positive charge introduced during the Al ion in diffusion at lower temperatures. Similar effects of bulk impurity redistribution in the vicinity of grain boundaries in solid electrolytes are often reported[2,9,12,20,36–38]. The built-in non-uniform dopant profile in the grain boundary vicinity due to segregation serves to influence the electrical behavior of the grain boundary following a restricted equilibrium scenario[39] which describes an intermediate scenario between the Mott-Schottky and Gouy-Chapman approximations (see section 10 SI for additional details). This case can only be evaluated via numerical simulations with knowledge of the dopant frozen-in profile in response to the thermal history – and is impractical for most experimental studies. We also noted that Si appeared to segregate at elevated temperatures (>1000°C), however its impact on our observed trends appears minimal. The fact that it is present in both sets of films (on $Al_2O_3$ and MgO) that still exhibit significant conductance differences at higher temperatures allows us to confidently rule out any significant role played by Si in our observations (see detailed discussion in SI section 9).

We summarize our findings in **Fig.** 5(a), that illustrates the increased degree of penetration along the GBs by Al and Mg ions by selective in-diffusion with increasing anneal temperatures, as well as increasing grain growth and segregation of the bulk Gd dopant with increasing temperature anneals for a given time, t. For temperatures of 600°C and below, partial selective penetration of substrate cations along the GBs is obtained. For intermediate temperatures of about 800°C, full relatively uniform penetration is achieved. At even higher temperatures on the order of 1000°C, grain growth occurs accompanied by Gd segregation to the grain boundaries as well as surface diffusion of the substrate derived cations on the surface of the film. Fig. 5(b) describes the relative impact of the segregation of the Gd bulk dopant, resulting in its accumulation in the space charge region and ultimately insertion of Gd into the core, causing a reduction of space charge potential and a decrease in space charge width (following a restricted equilibrium scenario), which either competes or supports the initial of Al and Mg on the grain boundary resistance. Clearly exemplified in this study is that by careful control of annealing time and temperature, it becomes possible to enrich the grain boundaries with selected elements, while minimizing the segregation of the bulk dopant.

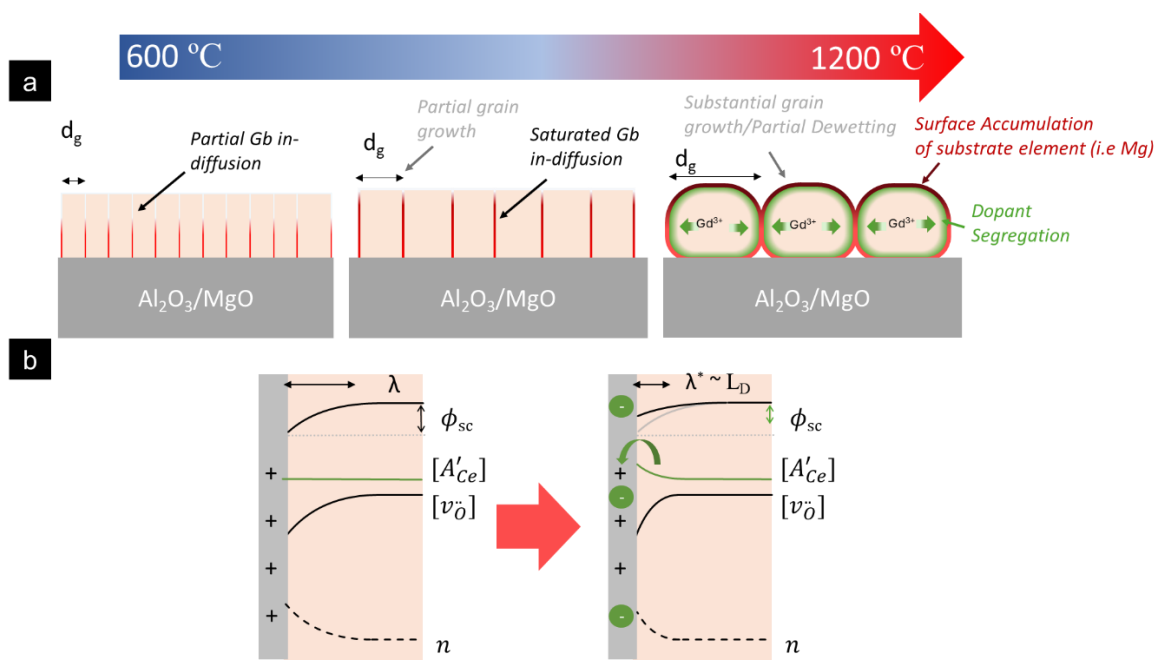

*Figure* 5 Physical model explaining observed trends associated with redistribution of cations at grain boundaries. 5a), from left to right: increasing anneal temperature leads initially to deeper GB penetration of dopants, ultimately leading at intermediate temperatures to saturation of in-diffused cations at the GBs and some grain growth. Further increases in annealing temperatures lead to substantial grain growth and Gd segregation from within the grains into the GBs. 5b) Spatial distributions of space charge potential, bulk dopant concentration, majority oxygen vacancy, and minority electron defect concentrations as $Gd'_{Ce}$ becomes sufficiently mobile at progressively higher temperatures (≥900°C) to segregate towards the grain boundaries, accumulating in the depletion region and contributing negative charge,[38] thereby reducing the barrier height $\phi_{sc}$ and the space charge width, $\lambda$, as displayed to the right. Full description provided in Fig S24.

In summary, we show that one can systematically decrease the grain boundary potential to near zero for films grown on MgO, allowing GDC films to exhibit conductivities characteristic of grain boundary-free, epitaxial-like samples. This has important implications for oxide based solid-state electrochemical devices since all such devices, regardless of conducting ion, rely on polycrystalline solid electrolytes or mixed ionic electronic conducting electrodes or membranes. Blocking of ion transport at grain boundaries is a common observation and, at a minimum, leads to increased ohmic losses and thus decreases device efficiency. Ion blockage at grain boundaries has furthermore been reported to contribute to device degradation, such as microcracking in solid oxide electrolysis cells[3] and dendrite formation in high-energy density sodium or lithium batteries[3,40].

We have also demonstrated that grain boundary barriers to ion migration can be substantially increased, e.g. from 0.25 V to 0.45 V, when annealing at temperatures < 900ºC on $Al_2O_3$ substrates. This leads to one of the largest activation energies ever reported for grain boundary resistance in the 3 at% GDC system (~1.6 eV). While to date, there have been few applications where large grain boundary barriers in solid electrolytes have been of interest, we recently demonstrated that such ion blocking barriers can be effectively utilized to detect UV and gamma rays[33,34]. As with other light or radiation detectors, a key figure of merit is a high dark resistance to achieve enhanced sensitivity. In the case of our ionic-based radiating detectors, this can be achieved not only by having large energy band gaps, but also by having grain boundary barriers as large as possible[41].

Open questions regarding the influence of grain boundary structure, i.e. grain boundary misorientation angles or complexions, on diffusivity of the elements and their solubility within the interface and their impact on the segregation of mobile charged species remain. For example, the highest space charge potentials that one can achieve in each system are likely to relate to the solubility limits of the dopants at these interfaces, as well as the competition between the relative segregation of the various charged mobile defects, bulk dopants, and grain boundary impurities to the boundaries[9,10]. How this might change as a function of grain boundary structure and grain boundary chemistry and how to identify and control this structure, and how the segregation processes depend on the existing presence of charged impurities will in turn be critical in defining upper limits for space charge values.

## 5. Summary, Outlook, and Conclusion

This study demonstrates the feasibility of systematically engineering grain boundary resistances in ion-conducting solid electrolytes over many orders of magnitude by careful incorporation of select cationic species at the interfaces. By employing targeted additives, it is possible to either minimize the built-in space charge potential to near zero or significantly increase it, depending on the application. As demonstrated in this work, a key consideration when attempting to increase the space charge barrier potential via selective addition of charged defect species is to limit bulk dopant redistribution by careful control of annealing temperatures, while it is a favorable effect in the case in which one wishes to decrease the barriers. This approach exemplifies the potential of extrinsic control of space charge properties through point defect engineering, offering parallels to the development of semiconductors where the ability to control their electrical properties by intentional doping revolutionized the field.

Relying on the in-diffusion of elemental impurities along grain boundaries at intermediate temperatures, a method explored in this work, enabled precise manipulation of space charge properties while maintaining the bulk properties unchanged. While our work focused on a single oxygen ion solid electrolyte (Gd-doped $CeO_2$), our findings are expected to have broad applicability to other solid state ionic systems, including protons, alkali metal (lithium & sodium), silver, and halide ion conducting materials. For example, this strategy can be extended to other ion-conducting systems of current interest for energy device applications, such as oxygen ion conducting yttria-stabilized zirconia (YSZ), lithium-ion conducting $Li_7La_3Zr_2O_{12}$ (LLZO), or proton-conducting $BaZr_{0.8}Y_{0.2}O_{3-\delta}$ (BZY). By employing lower annealing temperatures for extended durations, it becomes possible to minimize unwanted bulk dopant diffusion and microstructural evolution. However, critical challenges remain, such as understanding the influence of grain boundary structure—including misorientation angles and complexions—on dopant diffusivity, solubility, and space charge properties. These open questions will be crucial in determining the upper limits of achievable space charge potentials and in tailoring grain boundary properties for specific applications.

This work offers a clear demonstration of the ability to eliminate grain boundary resistive losses entirely, paving the way for reduced ohmic losses and improved performance in devices like electrolyzers and solid-state batteries. It also highlights the potential to intentionally increase barrier heights, enabling high dark resistances for novel applications such as radiation detectors with enhanced sensitivity. Nonetheless, further research is needed to explore how impurity ions substitute and diffuse in grain boundaries with different structures, the resulting charges associated with them, and the stability of space charge fields under elevated temperatures. These challenges present rich opportunities for experimentalists and theorists alike to advance this exciting field.

6. Methods

**Material Selection:** Gd doped ceria (GDC) was selected as a model material, as it is one of the most highly oxygen ion conducting solid electrolytes[42]. Grain boundary blocking of ion transport in this system has been well characterized and confirmed to be dominated by space charge depletion of oxygen vacancies[12,43]. A review of the theoretical basis describing the space charge properties of grain boundaries is provided succinctly in SI section 1 and covered in more detail in other publications[2,33,44]. Because of their relatively larger grain boundary resistance contributions, we chose a lower dopant level (3 at% Gd, in the following referred to as 3GDC) than what is typically selected to achieve maximum oxygen ion conductivity (10-20 at% Gd)[12].

**Substrate Selection:** MgO and Al$_2$O$_3$ substrates were selected based on several characteristics: 1) Insulating nature enabling electrical conductivity measurements of the overlaying films, 2) Binary composition enabling a simple up-diffusion process that is easier to interpret and 3) on the basis of the charge each of elements (Mg and Al) are expected to impart to the grain boundary core of CeO$_2$. Magnesium ions (Mg$^{2+}$), with a larger ionic radius, are expected to incorporate substitutionally onto Ce$^{4+}$ sites, thereby creating a doubly charged defect with net negative charge of 2-, which in conventional defect notation is given by $Mg''_{Ce}$. This is expected to compensate for the net positive core charge commonly found in ceria-based solid electrolyte grain boundaries[2,12,17,33], thereby lowering the built-in space charge potential. Conversely, the smaller aluminum ions (Al$^{3+}$), were anticipated to incorporate interstitially, creating defects with a net positive charge of 3+, which in conventional defect notation is given by $Al_i^{3\cdot}$, leading instead to an increase in space charge potential. Though strictly speaking substitutional and interstitial sites cannot be explicitly assigned in the grain boundary core due to ill-defined lattice structure, it still remains a helpful guide for evaluating the relative charge cations might possess within the interface.

**Pulse Laser Deposition Target**: The GDC powder used to prepare the Pulse Laser Deposition (PLD) target was synthesized through a co-precipitation route described by Spiridigliozzi et al[45]. A solution of gadolinium and cerium in stoichiometric proportion (3/97) was prepared by dissolving cerium nitrate, Ce(NO$_3$)$_3$:6H$_2$O, and gadolinium nitrate, Gd(NO$_3$)3:6H$_2$O (Strem chemicals) {99.99%}, in distilled water to reach a concentration of 0.1 molL$^{-1}$. The precipitation solution was prepared by dissolving ammonium carbonate in distilled water to reach a concentration of 0.5 mol$^{-1}$. The ammonium carbonate solution was calculated to have a molar excess of 2.5 compared to the total amount of cations in the nitrate solution. The ammonium carbonate solution was quickly poured into the vigorously stirred nitrate to trigger precipitation. The resulting residue was filtrated and washed four times in distilled water in a Buchner filter connected to a vacuum pump and dried at 100°C overnight, followed by calcination at 600°C for 1 h to obtain the GDC powder crystallized in the fluorite structure. The powder was then pressed into a 30 mm diameter disk with a uniaxial press (1 Tcm$^{-2}$) and sintered at 1500°C for 6 h, followed by cooling at 1°C min$^{-1}$ to limit crack formation, resulting in a pellet with 97% density.

**PLD thin film:** Both polycrystalline and nominally epitaxial samples were grown by PLD (Surface, Germany) using a KrF excimer laser with a 248 nm wavelength (Coherent, USA). The polycrystalline samples were directly grown on (001) MgO and (0001) Al$_2$O$_3$ single crystal substrates with a laser repetition rate of 10 Hz. The nominally epitaxial samples were grown following previously established protocols > 600ºC. Films were grown directly onto (0001) Al$_2$O$_3$ substrates[46], whereas a double buffer layer system of BaZrO$_3$ and SrTiO$_3$ (2-5 nm each) was initially grown onto (001) MgO substrate, to act as a seed layer[33,47]. After a base pressure of 7·10$^{-6}$ mbar was reached, pure oxygen was continuously leaked into the PLD chamber, keeping the total pressure at 0.013 mbar during film growth and cooling. During deposition of the polycrystalline films, the substrate temperature was held at 300°C (heating rate 10 °C/min and cooling rate 10 °C/min). The laser energy was set to 100 mJ, resulting in a power density of about 1 Jcm$^{-2}$. The target substrate distance was 7.5 cm. The resultant GDC films formed continuous pinhole-free layers. XRD was performed by a Bruker cobalt source D8 with a General Area Detector Diffraction System (GADDS). This system uses a conventional 1.6 kW sealed tube cobalt anode. Incident-side optics include a variety of double-pinhole collimators and mono-capillary devices, which are used to adjust the beam size, intensity, and divergence. The beam diameter for this instrument can range from 0.05 to 0.8 mm depending on the choice of collimator. The goniometer is a χ-cradle type, with full $\phi$ axis rotation and x-y-z translation. Including ω, this gives six positioning axes for the sample (not including the detector axis, 2θ). The Vantec-2000 detector has a very wide dynamic range and a maximum frame resolution of 2048 x 2048 pixels. Two-dimensional detectors facilitate the study of grainy and textured materials because the detector captures a slice of the Ewald diffraction sphere instead of a single point. Images of the cross-sections of the device were acquired at KAIST on a high-resolution scanning electron microscope (SEM, Hitachi SU8230). Samples prepared by cutting the substrate with a low-speed precision diamond cutter to ensure a perpendicular cutting angle.

**Annealing:** After depositing the samples, they were placed in $Al_2O_3$ crucibles and annealed in a $Al_2O_3$ tubular furnace at various temperatures (from 650 to 1250ºC) with 5°C/min ramp rates for 6 h in air to ensure complete oxidation of the lattice and to allow for microstructural evolution[48] and chemical diffusion. Inter-digitated ~7 mm long Pt or ITO electrodes were deposited on top of the film with the other dimensions outlined in table 1 of the Supplementary. The Pt electrodes were deposited at room temperature by DC sputtering while ITO was deposited by PLD using similar conditions to those for the GDC films. By employing substrate-assisted in-diffusion rather than top-down metallic sources[32], we preserve film integrity and minimize extraneous effects on conductivity measurements.

**Electrochemical Impedance Spectroscopy (EIS):** All electrochemical measurements were performed in a Linkam stage HS600 with a quartz window, allowing for heating and illumination of the sample. The chamber was flooded with 50 sccm of synthetic air to ensure oxidizing conditions throughout the measurements. All EIS measurements were performed with an MFIA Impedance Analyzer (Zurich Instruments). A 100mV amplitude was necessary for a sufficiently high current response in the highly resistive samples. The frequency range was from 0.01 Hz to 1 MHz. Each EIS spectra was repeated at least once.

**SEM and TEM characterization:** SEM characterization was performed using a Zeiss Merlin SEM. Samples were carbon-coated prior to imaging. Grain size measurements were performed using the line intercept method, with at least 100 intercepts at each temperature. Grain growth activation energies were found by assuming ideal grain growth (grain growth exponent = 0.5), and an initial grain size of 30 nm. TEM samples were prepared using two methods: cross-section wedge polishing and FIB liftout. In either case, final polishing was performed by low-energy broad-beam $Ar^+$ ion milling. STEM characterization was performed on a Thermo Fisher Scientific Themis Z S/TEM at 200 keV. EDS quantification was performed using the Velox software. For line traces, a 5 px mean pre-filter was applied prior to quantification. For spectrum image maps, a 5 px standard deviation Gaussian pre-filter was applied. Empirical background subtraction was performed prior to quantification.

**In diffusion measurements by SIMS:** Diffusion profiles of Mg and Al following annealing of the 3 at% Gd doped ceria films grown on MgO and $Al_2O_3$ substrates respectively were obtained through SIMS analysis measured by TOF-SIMS5 instrument (ION-TOF GmbH, Germany). Positive detection mode was used for the detection of $Ce^+$, $Gd^+$, $Al^+$, and $Mg^+$. Continuous bombardment of focused $Bi^+$ ion beam was made over a 100 μm × 100 μm region during the measurement with a 30 keV Bi ion gun. Released secondary ions were analyzed using a time-of-flight-detector. The sputter beam was set at 2 keV for $O_2^+$ ion-sputtering made over a region of 300 μm × 300 μm throughout the analysis. The analysis was conducted with a cycle time of 100 μs.

# Acknowledgements

TD, CG, JL and HT acknowledge support from the U.S. Department of Homeland Security, Countering Weapons of Mass Destruction Office, under awarded grant 22CWDARI00046. This support does not constitute an express or implied endorsement on the part of the Government. W. Jung and Y.B. Kim acknowledge support from the National Research Foundation of Korea (NRF) grant funded by the Korea government (MSIT) (RS-2024-00452853) and Research Institute of Advanced Materials (RIAM). This work was carried out in part using the MITnano characterization facilities.

# Author Contributions

T.D suggested the original idea. T.D and H.L.T designed the experimental protocol. T.D performed the sample preparation and electrical conductivity measurements. Y. B. K performed the SIMS and XRD measurements. C.G. performed the SEM studies and prepared the lamellae for TEM alongside performing the TEM measurements with J.L. providing guidance in interpretation of results. H.L.T. supervised the work and provided guidance throughout the project. T.D and H.L.T wrote the paper with assistance of Y. B. K and C.G. All the co-authors discussed the results and helped to revise the manuscript

Supplementary Information

# Grain Boundary Space Charge Engineering of Solid Oxide Electrolytes: Model Thin Film Study

T. Defferriere[1*], Y.B. Kim[2,3], C. Gilgenbach[1], J. M. LeBeau[1], W. Jung[2,3], H.L. Tuller[1*]

[1]Department of Material Science and Engineering, Massachusetts Institute of Technology, Cambridge, MA 02139, USA
[2]Department of Materials Science and Engineering, Seoul National University, Seoul 08826, Republic of Korea
[3]Research Institute of Advanced Materials, Seoul National University, Seoul 08826, Republic of Korea

*to whom correspondence should be addressed: tdefferr@mit.edu, tuller@mit.edu

Contents

## 1. Space Charge Model

The conductivity of a grain boundary across the space charge zone can be defined as:

$$\sigma_{gb}(x) = \sigma_{bulk} e^{-\frac{ze\Delta\phi(x)}{k_B T}} \tag{S1}$$

which describes how the space charge zone conductivity relates to the bulk conductivity ($\sigma_{bulk}$) and the depletion of ionic carriers within the depletion zone following the spatial distribution along x (described by space charge width - λ), of the space charge potential in the vicinity of the interface. Eq. S1 shows that as the space charge potential approaches zero, the grain boundary conductivity approaches the bulk conductivity.

The potential distribution is found by solving Poisson's equation, which describes how the system responds to the presence of a net core charge $Q_{core}$, but no exact analytical solution can be obtained[1]. One can, however, derive an approximate expression in the case of sufficiently large space charge potentials (i.e., $\frac{2ze\Delta\phi(0)}{k_B T}>3$). For the sake of simplicity, we can also consider a simplified dilute limit picture and apply the Mott-Schottky case to describe the space charge zone, whereby we assume the dopant profile to be flat and frozen-in during processing, and only the oxygen vacancies can redistribute within the proximity of the grain boundary. As previously derived by several authors[1–3] in this simplified scenario, the grain boundary resistance and its relation to the space charge potential in the grain boundary core ($\Delta\phi(0)$) can be obtained by integrating Eq. S1, yielding:

$$\frac{R_{gb,tot}}{R_{bulk,tot}} \approx \frac{e^{\frac{ze\Delta\phi(0)}{k_B T}}}{\frac{2ze\Delta\phi(0)}{k_B T}} \tag{S2}$$

The effective grain boundary space charge width, defined as $\delta_{GB} = 2\lambda$, generally thought of as the electrical grain boundary width, where we typically neglect the grain boundary core contributions is defined according to λ, being the Mott-Schottky space charge width:

$$\lambda = 2\left(\frac{kT\varepsilon}{2e^2[Gd'_{Ce,bulk}]}\right)^{0.5}\left(\frac{e\Delta\phi(0)}{kT}\right)^{0.5} \tag{S3}$$

which relates to the Debye length $L_D$ (first term in Eq. S3 in parenthesis) and the space charge potential. The space charge potential, in turn, is defined according to the balance between the core charge at the grain boundary and the amount of opposite compensating charges in the space charge zone. In the Mott Schottky case, where the bulk dopant is assumed to have a flat concentration profile, the space charge potential in the core can be equated to:

$$\Delta\phi(0) = -\frac{Q_{core}^2}{8e \cdot \varepsilon \cdot [Gd'_{Ce}]} \tag{S4}$$

where $Q_{core}$ (Core Charge in C/cm²) and $\varepsilon$ The dielectric constant. This equation shows how the potential barrier heights stem from a charge balance strongly dictated by the core charge ($Q_{core}$) and secondarily influenced by the bulk dopant concentration. This equation explains both the effectiveness of our chemical modification of the space charge potential, owing to the squared term in the numerator, and why we decided on a lightly doped Gd doping level (3at%) as the concentrations of the dopant. $[Gd'_{Ce}]$ is in the dominator, with smaller values leading to larger space charge potentials.

It is also possible to derive an approximate solution for another scenario, called the Gouy-Chapman case, that considers the dopant to also be mobile and able to redistribute in the space charge zone throughout the measurement conditions. In that case, the space charge width is smaller and equates to the Debye length (i.e $\lambda = L_D = \left(\frac{kT\varepsilon}{2e^2[Gd'_{Ce,bulk}]}\right)^{0.5}$). The Gouy-Chapman case, however, is only valid when the dopant is mobile during the measurement conditions, which is only valid at elevated temperatures (> 1000°C) and is generally not considered for lower temperature conductivity measurements, where the dopant is effectively immobile during the measurement time, warranting the use of the Mott-Schottky approximation[4]. We recognize that this equation is a first-order approximation that can lead to errors in absolute values, as non-uniform concentrations of the dopant profile in the grain boundary vicinity arise during high-temperature processing conditions and end up subsequently frozen-in upon cooling (>30%)[4]. This leads to a so-called *restricted equilibrium* scenario[4] (See Fig. 5 (b) in the main manuscript and Fig S24 at the end of the SI), which is out of equilibrium and lies between the Mott-Schottky and Gouy-Chapman cases. This can only be evaluated via numerical simulations with knowledge of the dopant frozen-in profile, impractical for most experimental studies.

Furthermore, for our studies, we use epitaxial films as a measure of the bulk resistance to fit the space charge potential, assuming it does not substantially change with the annealing protocol through the sintering temperature window. We determine this to be acceptable in our case as we are more interested in general trends, and the results obtained from the resistivity ratio can be self-consistently compared.

It is also worth noting that while most studies of GB resistivity in ionic conductors derive a single effective barrier potential, GBs in given solids are known to differ depending on their misorientation angles between adjacent grains, impacting impurity and defect segregation, and leading ultimately to a range of GB potentials[5] that cannot be easily distinguished from simple electrical measurements. This is exacerbated by the fact that electrical measurement techniques measure the weighted average of all current paths across the numerous grain boundaries, dominated by the paths of least resistance[4]. This implies that approximate analytical solutions for fitting grain boundaries space charge potentials based on electrical measurements are mainly used to consider general trends and for the sake of discussion, as we do in this study.

## 2. Electrode Geometry

*Table* S1 List of electrode dimensions deposited on GDC films grown on $Al_2O_3$ and MgO and annealed at various temperatures

| Temperatures (ºC) | Wide (μm) | Digits | Spacing (μm) | Material |
|---|---|---|---|---|
| **MgO** | | | | |
| 850 | 350 | 6 | 200 | Pt |
| 950 | 350 | 6 | 200 | Pt |
| 1050 | 350 | 6 | 200 | Pt |
| 1150 | 350 | 5 | 200 | Pt |
| 1250 | 250 | 5 | 200 | Pt |
| Epi | 350 | 6 | 200 | Pt |
| **$Al_2O_3$** | | | | |
| 750 | 250 | 8 | 350 | ITO |
| 850 | 250 | 8 | 350 | ITO |
| 950 | 250 | 5 | 350 | ITO |
| 1050 | 250 | 6 | 350 | ITO |
| 1150 | 250 | 5 | 350 | ITO |
| Epi | 250 | 6 | 350 | ITO |

## 3. Structural Characterization

The X-ray diffraction patterns of both GDC polycrystalline films grown on $Al_2O_3$ and MgO are consistent with the expected cubic fluorite structure exhibiting grains with (111) and (200) dominant orientations (see Figure S1c). The XRD patterns of the underlying substrate, $Al_2O_3$, and MgO are present at 41.5º and 42.9º 2θ.

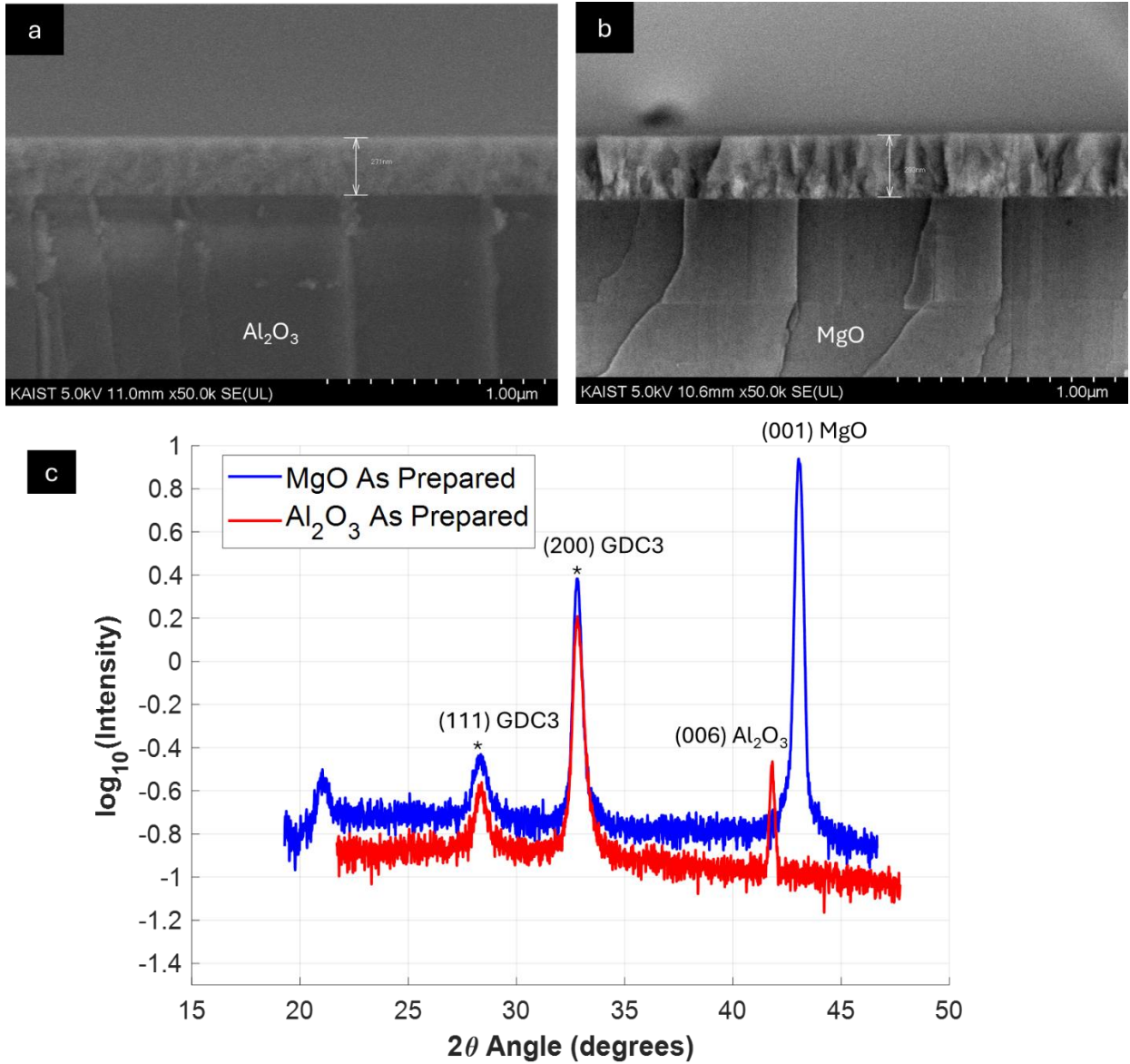

*Figure S1:* (a,b) SEM cross-sectional micrographs, and c) XRD patterns of GDC films as grown (~250°C) on $Al_2O_3$ and MgO respectively.

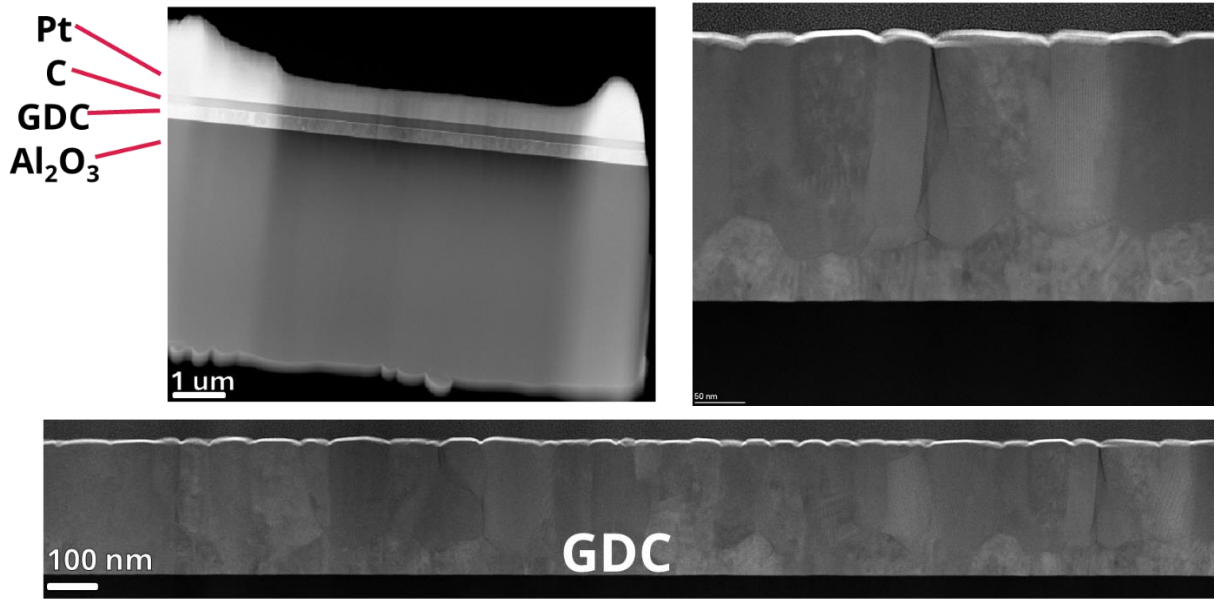

*Figure S2: Characterization of GDC film grown on $Al_2O_3$ by SEM (upper left image) and TEM (upper right image and lower image) shows it to be polycrystalline with columnar grains.*

## 4. Structural Evolution of GDC3 films on MgO as a function of temperature (XRD/SEM)

We display the 2 theta scans for the GDC films grown on $Al_2O_3$ substrates and subsequently annealed from 650ºC up to 1250ºC in **Fig.** S3(a,b). Similar to the as-grown films, no evidence of secondary phases is observed. This is additionally corroborated from STEM images shown in **section 7** of select boundaries at various key temperatures, where no evidence of secondary phases can be observed. While no clear trend in change in 2 theta position is observed for the dominant film peaks, with increasing annealing temperature, the peak height of the (111) decreases relative to the (200) peak orientation, indicating a gradual increase in preferential film orientation texture towards the (200) orientation, which is fully achieved by 1000°C, at which point the (111) peak is no longer visible. Moreover, when looking at the rocking curves displayed in **Fig.** S3(c), performed on the dominant (200) peak, we observe that the peak shows an apparent decrease in its full-width half max, implying a decrease in wide angle in favor of lower angle grain boundaries, consistent with grain growth and improvement in film quality. We observe similar trends for films grown on MgO and annealed from 650ºC to 1250ºC, as displayed in **Fig.** S4, though the disappearance of (111) seems to occur at lower temperatures (~750°C).

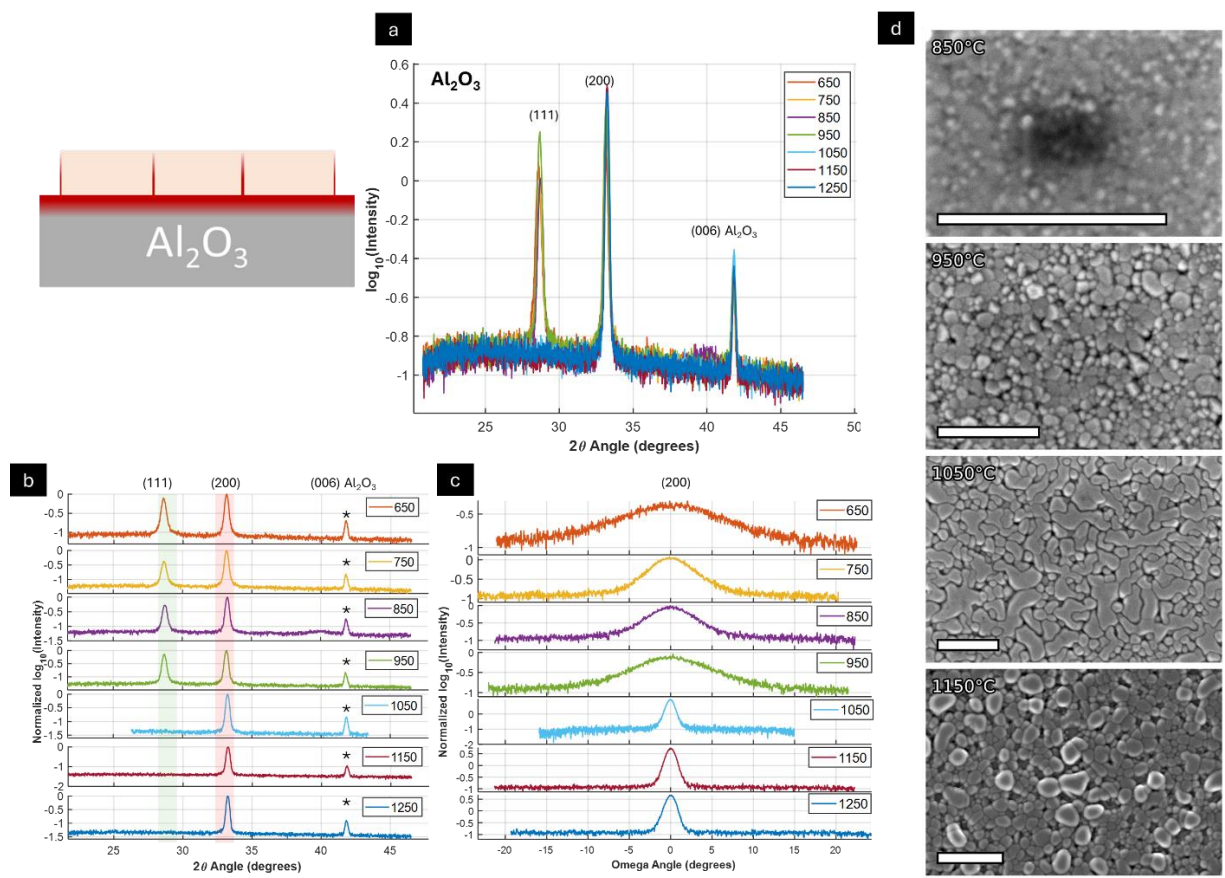

*Figure S3:* a) Overlaid view of XRD spectra for GDC films grown on $Al_2O_3$ and annealed from 650ºC to 1250ºC. b) Normalized cascaded view of the XRD patterns for each film c) Rocking curve of the (111) and (200) orientation as a function of annealing temperature. d) Micrographs of GDC thin films grown on $Al_2O_3$ and annealed at different temperatures. Images were collected at 3 keV electron energy, working distance 4 mm, using a secondary electron detector. Scale bars 1 um.

Scanning electron microscope (SEM) measurements of the surface of the films were performed to characterize the microstructure and systematically measure the grain size of the synthesized films. In **Fig**. S3(d), representative SEM micrographs are shown for GDC films grown on $Al_2O_3$ substrates, while similar plots are present in **Fig.** S4(d) for GDC films grown on MgO substrates. For both cases, similar microstructural evolution was observed. As deposited, and at low annealing temperatures, films are equiaxed with fine grain structure (d≈30 nm). A columnar grain structure is observed at intermediate temperatures, e.g., 950°C. Additionally, abnormal grain growth occurs, likely due to significant surface energy differences between grains of preferred orientation and grains of similar orientation merging during growth. At extreme temperatures, e.g., 1150°C, the average grain size becomes much greater than the film thickness, and dewetting occurs, leaving roughly spheroidized grains.

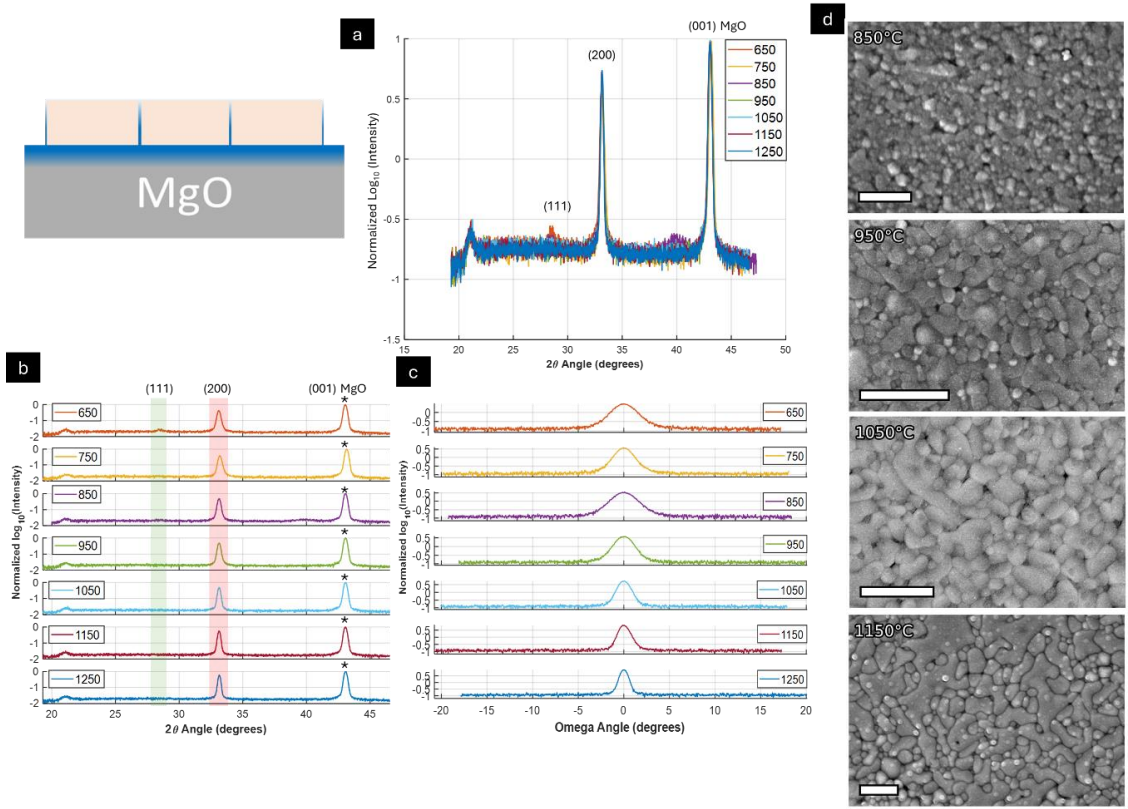

*Figure S4:* a) Overlaid view of XRD spectra for 3GDC films grown on MgO and annealed from 650°C to 1250°C. b) Normalized Cascaded view of the XRD patterns for each film c) Rocking curve of the (111) and (200) orientations as a function of anneal temperature. d) Micrographs of GDC thin films grown on MgO and annealed at different temperatures. All images were collected at 15 keV electron energy, working distance 12 mm. The 1075°C micrograph was taken with a backscatter detector; the others were taken with a secondary electron detector. Scale bars 1 um.

By applying an Arrhenius fit to the grain sizes measured at different temperatures, an activation energy for grain growth of 1.55 eV ($Al_2O_3$) and 1.32 eV (MgO) is found in **Fig**. S5. Details of the grain size measurements and fitting process are provided in the methods section. Similar microstructural evolution trends and activation energies in nanocrystalline ceria thin films have previously been reported by Rupp et al[6] and Esposito et al[7], when columnar thin films of 26 at% and 10 at% Gd doped $CeO_2$ were grown on $Al_2O_3$ substrates by spray pyrolysis and PLD. These exhibited mixed polycrystalline films characterized by (111)/(100) orientations, when exposed to increased thermal anneals (from 600 to 1200°C), ultimately also led to single grain orientation. However, in those cases, the (111) orientation ended up dominating. We note that contrary to our initial films, they observed a predominance in (111) orientation in the as-deposited state, whereas we observe a predominance of (100), which suggests that the end results may arise from a competition in grain boundary interface mobility. Presumably, the (100)/(111) may be the slowest, and our final results depend on the initial distribution of boundary orientations. Additional differences in grain boundary mobility might arise from the in-diffusion of Al vs Mg, but generally, such observations are unimportant to this study.

Further evidence that the trends in changing microstructure are not causing the major observations in conductivity observed in our study stems from the case of the GDC3 film grown on $Al_2O_3$ annealed at 1150°C, where an unexpected decrease in conductance relative to the sample annealed at 1050°C is observed, as displayed in Fig. 2(b). This occurs even though the activation energy continues to decrease. We believe that the change in conductance coincides with the decrease in grain size that can be observed in Fig. S3(d) and Fig. S5(a). We suspect that the abnormal morphology observed for the sample annealed at 1050°C may have reached a point of instability and broken up due to surface tension. The decreased grain size would in turn correlate with an increase in total number of grain boundary barriers along the current path adding additional series resistance, consistent with the decrease in conductance. This supports the idea that the change in activation energy is not caused by changes in microstructure. The mechanism leading to abnormal changes in grain morphology is outside of the scope of this study as the activation energy trend reported in Fig. 3 remains consistent with the thermal annealing trend.

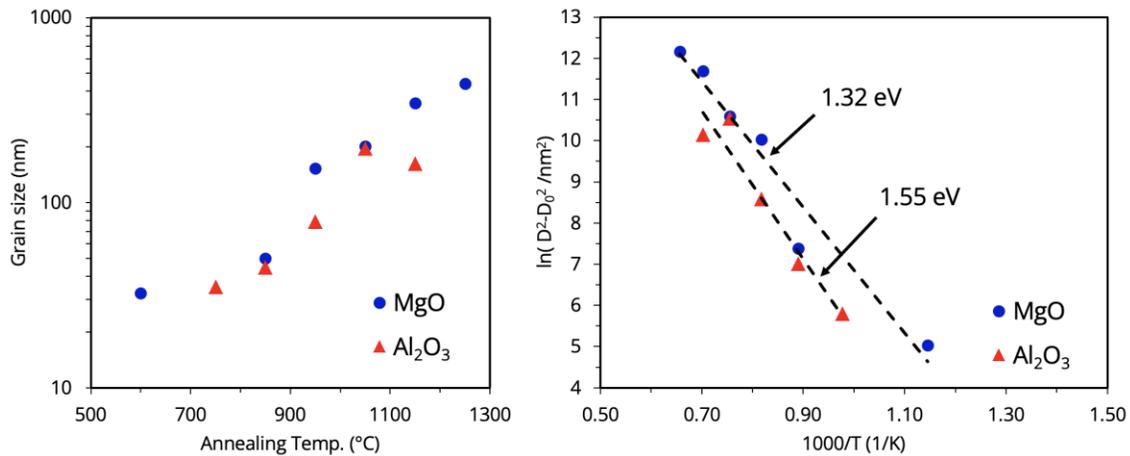

*Figure S5: a) Measured grain sizes as a function of annealing temperature for GDC thin films grown on MgO (blue circles) and $Al_2O_3$ (red triangles). b) An Arrhenius plot on the right shows the thermal activation of grain growth. An ideal grain growth exponent of 0.5 and an initial grain size of 30 nm were assumed. An activation energy of 1.32 eV is found for MgO, and 1.55 eV for $Al_2O_3$.*

## 5. Impedance spectra for all GDC3 films on MgO

Exemplary measured electrochemical Impedance Spectra (EIS) for all thin films grown on $Al_2O_3$ and MgO measured at 350°C are shown in **Fig.** S6(a,b). Only a single large semicircle can be observed at higher frequencies, with a small tail at lower frequencies, consistent with our previous measurements on similar thin film samples[3]. The main observed semicircle for GDC is ascribed to the combination of bulk and grain boundary contributions. The bulk contribution is not visible because it is shielded by the stray capacity that forms between the substrate and the electrodes[8,9]. The low-frequency arc is expected to be associated with the electrode contribution, which is not the focus of this report and is therefore excluded from subsequent analysis. The characteristic frequency and time constant $\tau\ (=2\pi f_{char})^{-1}$ of the main semicircle depends mainly on the sample's grain boundary resistance and electrode geometry. Note for polycrystalline GDC3 films grown on $Al_2O_3$, the lowest resistance measured at 350°C, observed in the inset of Fig S6, is equal to ~$4\times10^6$ Ω, which is $10^3$ times more resistive than the most resistive film annealed at 850°C, possessing a value of ~$7\times10^9$ Ω. The epitaxial value is even smaller and is only barely visible in the inset near the origin, with a resistance value of ~$0.5\times10^6$ Ω. On the other hand, for the case of films on MgO, the lowest values are recorded for films annealed at 1050°C, equal to ~ $0.7\times10^6$ Ω, vs the highest resistance for the film grown at 850°C with a value of $1\times10^8$ Ω.

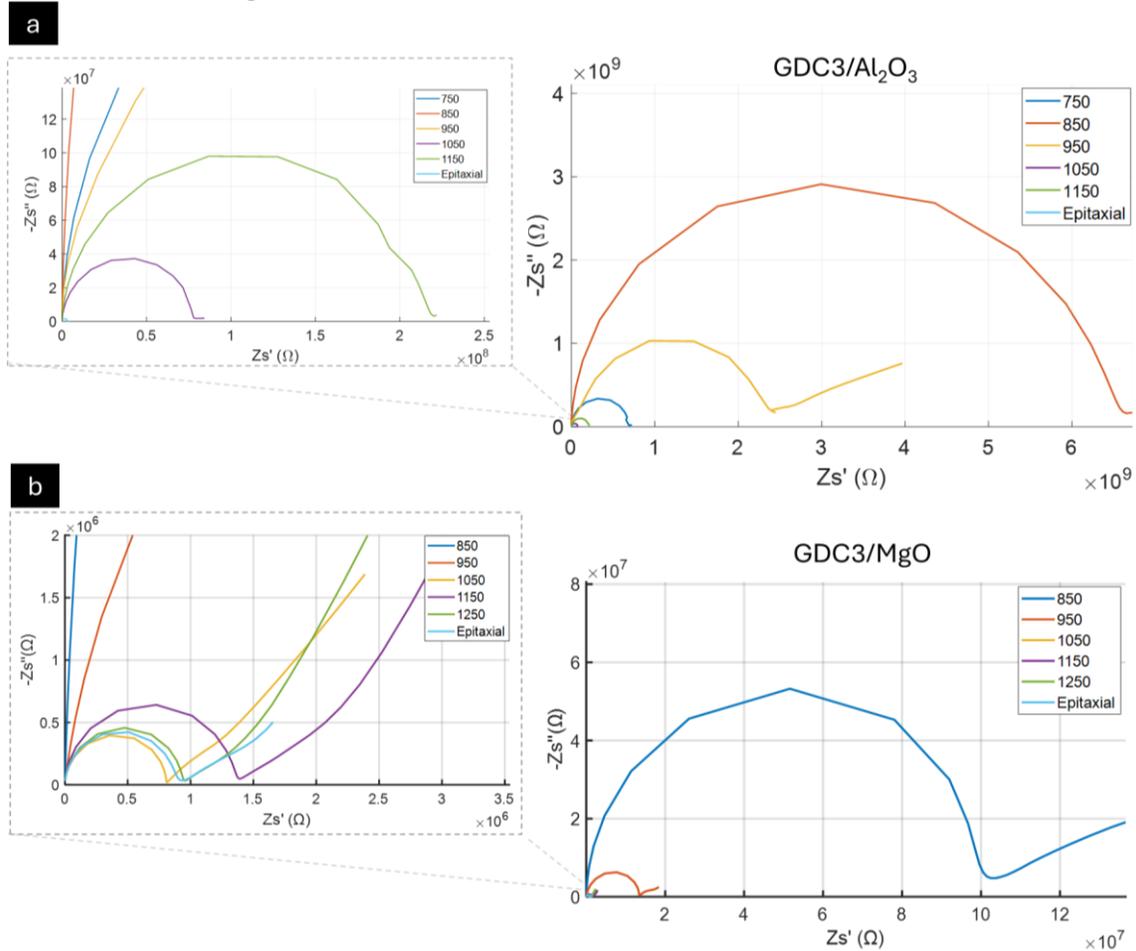

***Figure** S6: (a,b) Example Nyquist plots of the complex impedance response obtained at 350°C under open-circuit conditions for a GDC3 films on $Al_2O_3$ and MgO substrates on right, respectively. Smaller plots to left represents a zoom-in of data near the origin to showcase the low resistance impedances at the more elevated temperatures near the origin at the highest frequencies.*

## 6. Epitaxial behavior MgO vs Al$_2$O$_3$

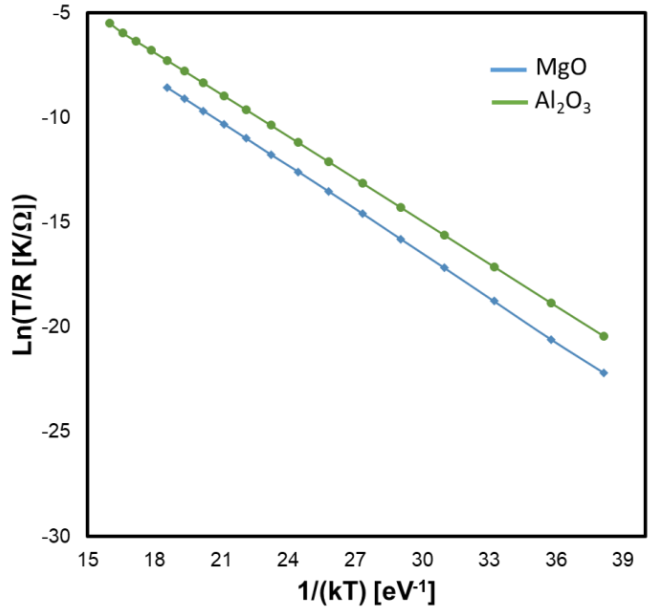

***Figure*** *S7: Semi-log plots of Arrhenius dependence of conductance G as (1/R) x T vs. 1/kT. Data points derived from impedance plots obtained for epitaxial films grown on MgO and Al$_2$O$_3$.*

As shown in **Fig.** S7, the nominally epitaxial films grown on MgO and Al$_2$O$_3$ show similar activation energies but deviate in absolute resistance by a factor of 5. Variation in measured conductance may arise from differences in film thickness, electrode geometry, digit spacing, and built-in strain owing to the differing near epitaxial growths on the two structurally different substrates. For the purpose of our study, understanding the exact sources of these differences is not of importance as these near epitaxial films solely serve as bulk conductance references against those of the respective polycrystalline films.

## 7. Impedance Fitting and Space Charge Model Fitting

*Capacitance Values:* The spectra are fit using distributed elements composed of a R//CPE, where CPE is a constant phase element from which capacitance values were calculated based on the expression $C=(R^{(1-n)} Q)^{(1/n)}$, where Q is the constant phase element capacitance, R the resistance and n the non-ideality factor obtained from the fitting procedure. The measured capacitances obtained for all films, including the epitaxial ones, are all about equal to $\sim 1\text{-}5\times 10^{-12}$ F. Using table S1, we can calculate the expected substrates capacitance assuming a simple parallel plate capacitor, described by: $C = \varepsilon_r \varepsilon_0 \left(\frac{LN}{d}\right)$, where L is the length of digits, N the number of pairs of digits, d the distance between digits, and $\varepsilon_r \varepsilon_0$ the dielectric constant of the materials. From this relation, we obtain capacitance values ranging from $7\times 10^{-13}$ to $5\times 10^{-12}$ F, which align well with the measured values.

*Space charge Model Fitting:* As discussed in section 1 of SI, we can use equation S2 to fit our impedance results to obtain an approximate value for the space charge potentials of each film. As discussed in section 5 of SI, at higher frequencies we only observe a single large semicircle in each impedance spectra for all films, which is ascribed to the combination of bulk and grain boundary contributions. The bulk contribution is not visible because it is shielded by the stray capacity that forms between substrate and electrodes. To use equation S2, we therefore use the impedance value of the epitaxial samples as a first order approximation of the bulk resistance. We believe this to be a valid approximation as the bulk resistance of the polycrystalline films is not expected to vary with annealing temperature (see section S4 indicating that there is no obvious change in strain state of the film that would induce a change in bulk conductance). Fitting procedure and results are displayed in **Fig**. S8 (b,c) below.

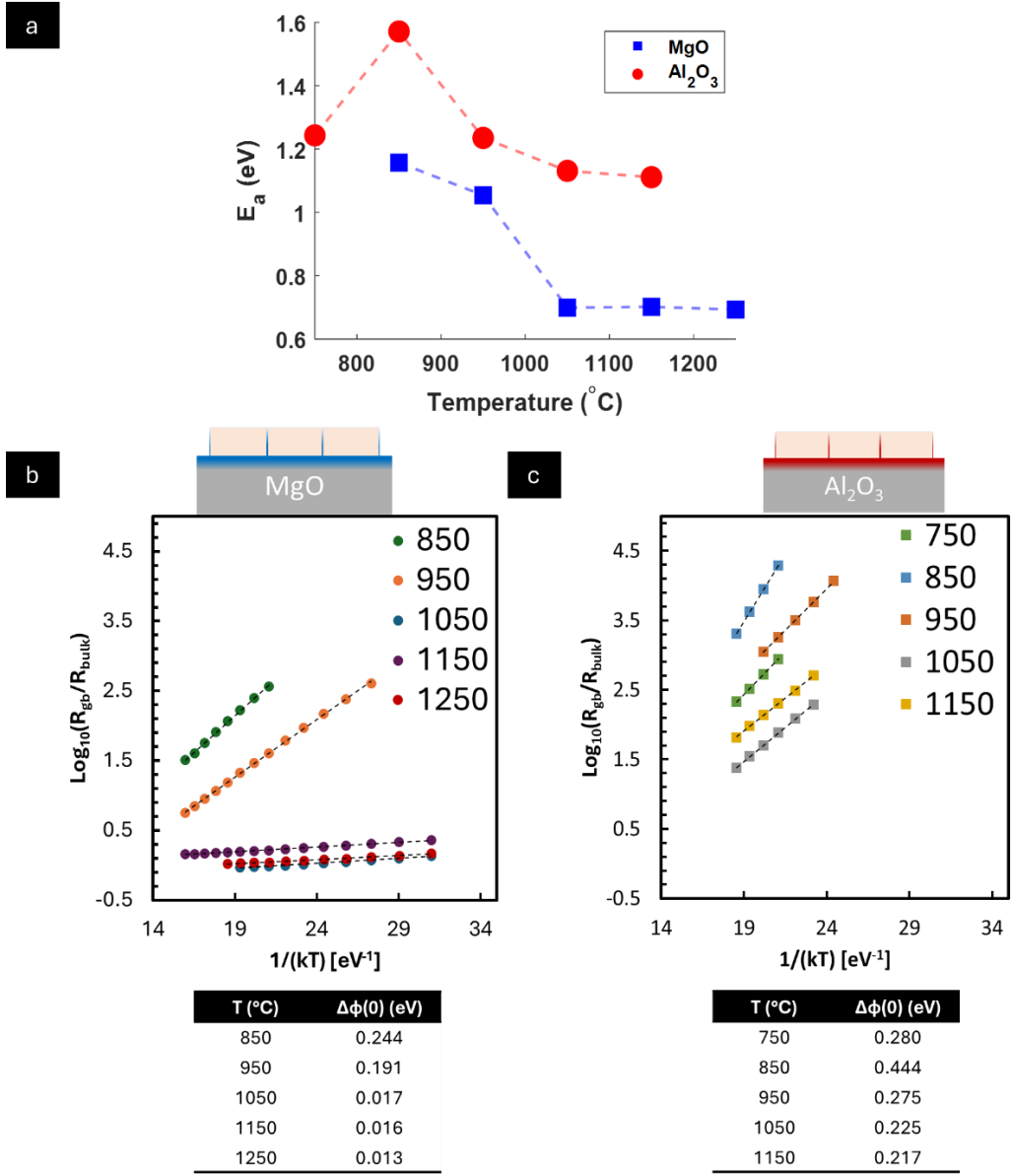

***Figure*** *S8: a) Total activation energy vs annealing temperature for films grown on MgO vs $Al_2O_3$ (b,c) Plots of $Log_{10}\left(\frac{R_{gb}}{R_{bulk}}\right)$ vs $1/kT$ used for the fitting space charge potentials with the aid of equation S2 for films grown on MgO (left) and $Al_2O_3$ (right). $R_{bulk} \sim R_{epitaxial}$ assumed for the fitting procedure to calculate the space charge potentials $\Delta\phi(0)$ listed as a function of temperature in the tables below each plot.*

# 8. Secondary Ions Mass Spectroscopy (SIMS) Data

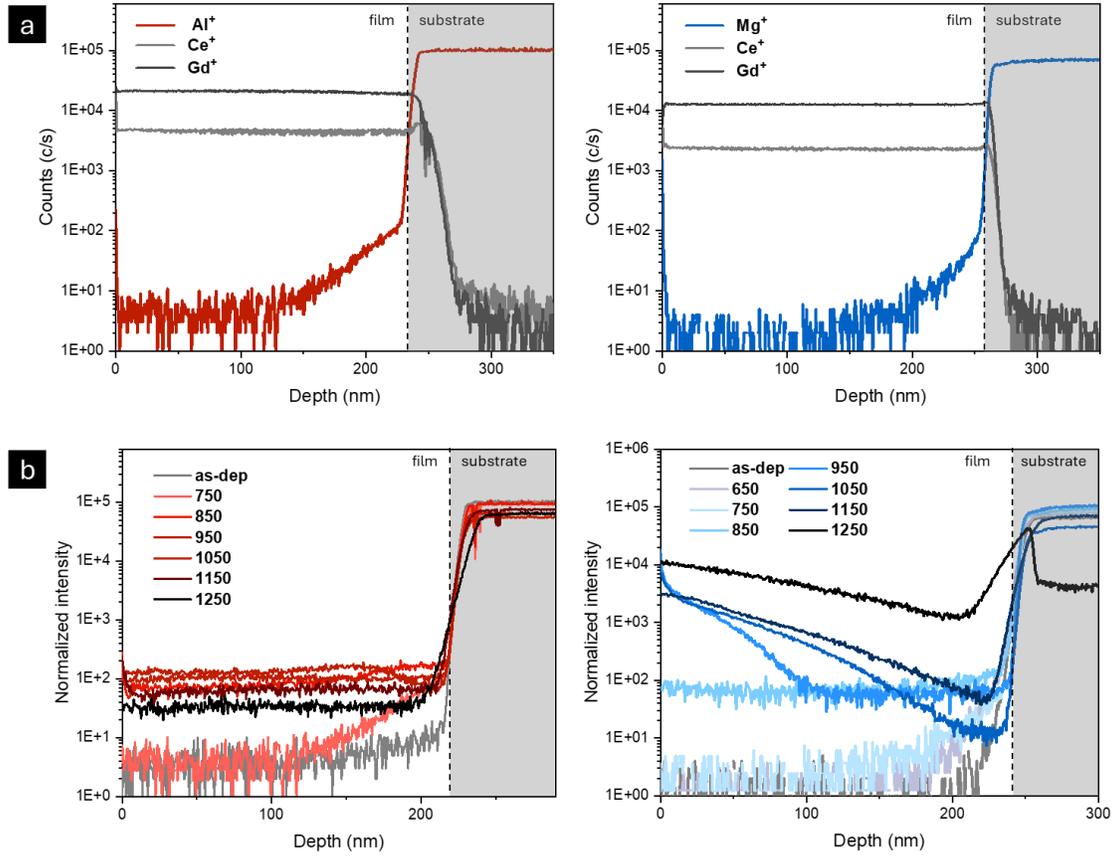

***Figure*** *S9: (a) ToF-SIMS depth profile of Al+, Ce+, Gd+ and Mg+, Ce+, Gd+ in GDC deposited on Al$_2$O$_3$ (left) and MgO (right) after annealing at 750 °C for 6 hours. The dotted line signifies the interface between the GDC film and the substrate. (b) Depth profiles of Al+ and Mg+ as in (a) for various listed annealing temperatures. Detailed view of the SIMS profiles for all samples provided in **Fig.** S10 and S11.*

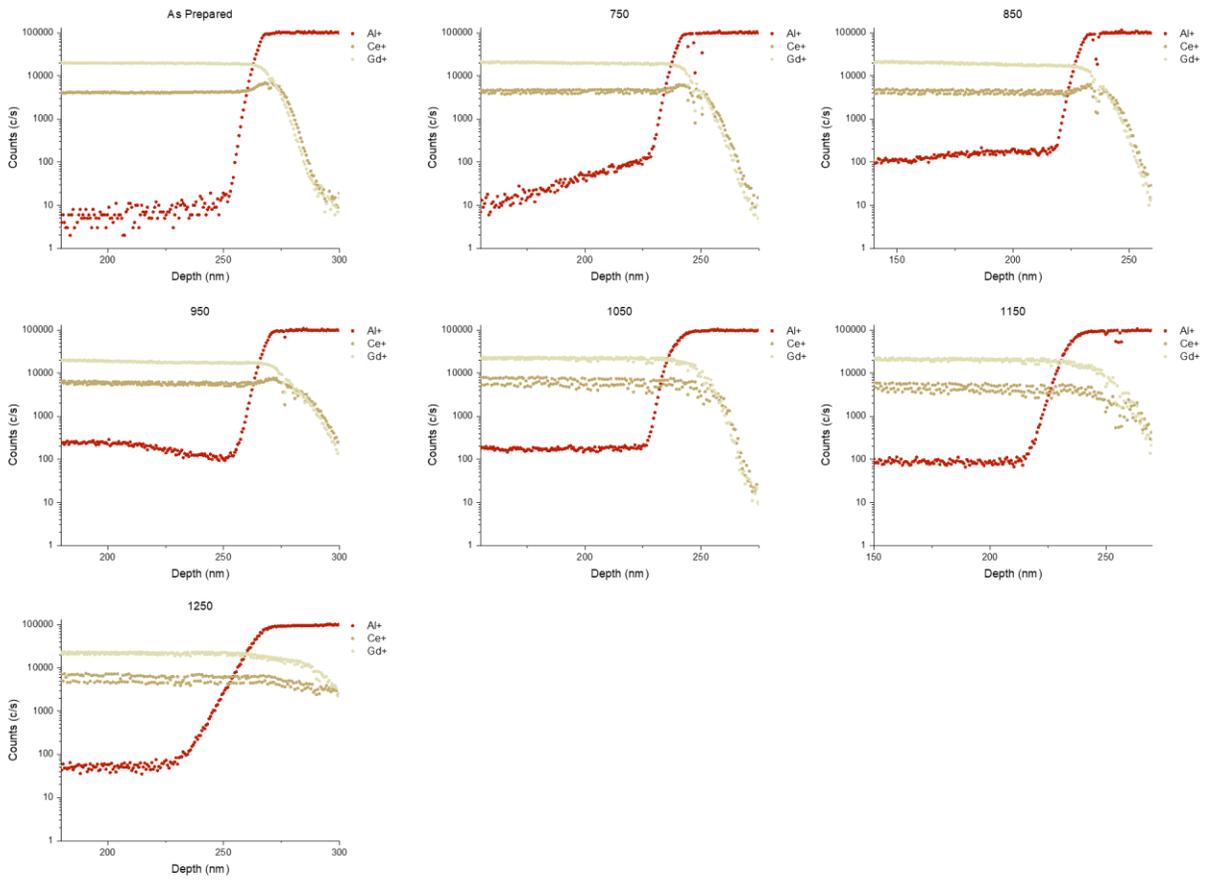

***Figure*** *S10: ToF-SIMS spectra profiles of cations for GDC deposited on $Al_2O_3$. Depth profiles were obtained for as-deposited samples and samples annealed at varied temperatures for 6 hours.*

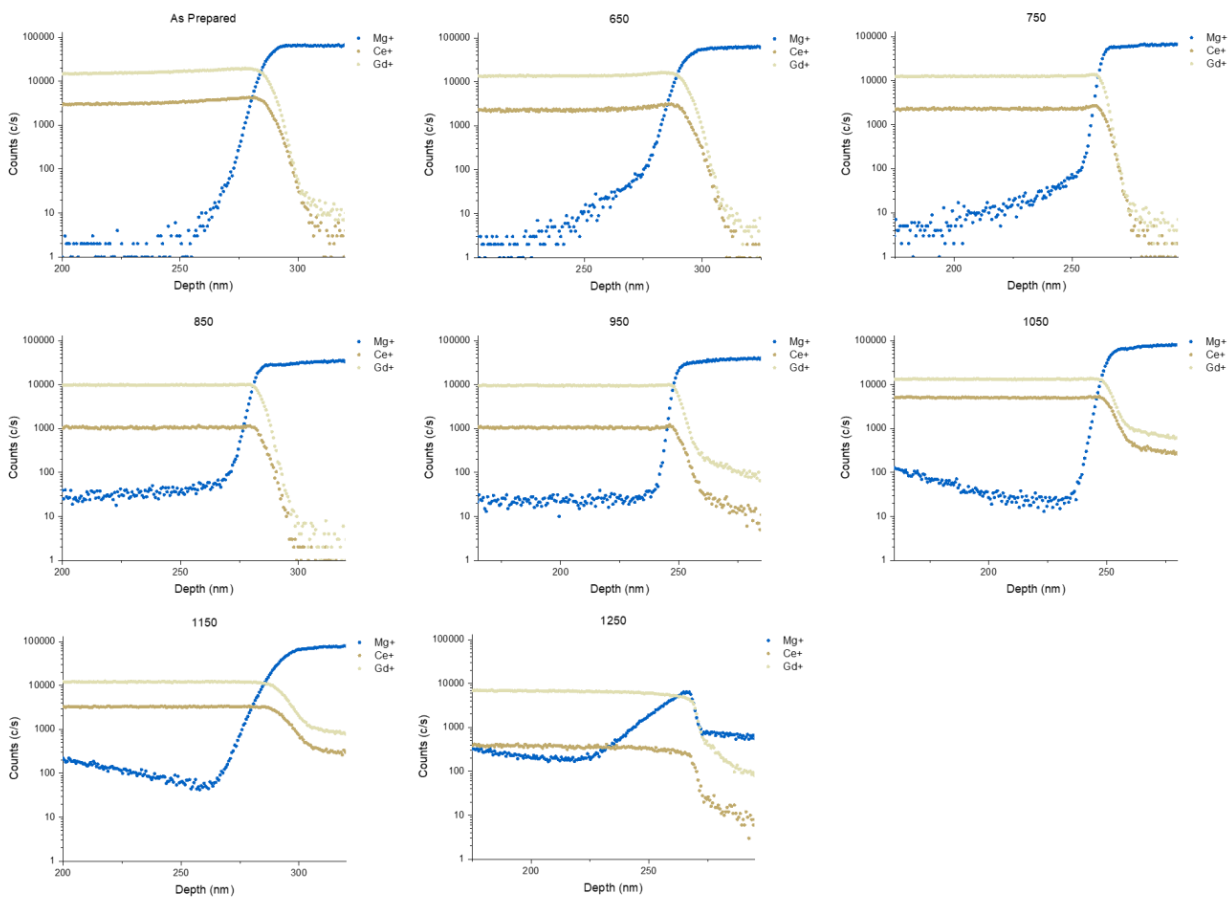

*Figure S11: ToF-SIMS spectra profiles of cations for GDC deposited on MgO. Depth profiles were obtained for as-deposited samples and samples annealed at varied temperatures for 6 hours.*

As displayed in **Fig.** S9, for both Al and Mg, the concentration of dopants within the GBs exhibits a saturated behavior when annealed at 850°C for 6 h, where the concentration of in-diffused species presumably reached the solubility limit within the grain boundaries of the deposited GDC films. For samples deposited on $Al_2O_3$, the saturation concentration slightly decreased with increasing annealing temperatures above 950°C. This can be rationalized by noting the reduction in grain boundary concentration due to grain growth during high-temperature heat treatments that eventually lowered the apparent grain boundary solubility per surface area. Meanwhile, samples deposited on MgO showed increased Mg concentration at the surface following annealing at temperatures above 950°C. The dopant concentration increased continuously with increases in annealing temperature, suggesting additional diffusion occurring along the film surface, potentially exacerbated by the increased amount of surface area due to grooving and dewetting of the film, as described in previous reports.[10]

For annealing profiles below 950°C, grain boundary diffusion coefficients were fitted by applying a 1D diffusion equation (Equation S5) to the dopant depth profiles, which were normalized (see **Fig.** S12 and S13) with respect to the host $Ce^+$ signals.

$$I(x) = A \cdot \text{erfc}\left(\frac{x}{2\sqrt{D_{GB} \cdot 21600}}\right) + L \qquad (S5)$$

Here, I(x) is the normalized intensity at depth x, A in a proportional constant, and $D_{GB}$ is the diffusion coefficient of the dopant element. We adopted constant source grain boundary conditions for the dopants, and the background signal (L) was set by the dopant concentration in the as-deposited samples. The diffusion behavior of dopants within the grains was also investigated by fitting the near-interface region depth profiles above 1000°C (see **Fig.** S14 and S15) using a modified solution for diffusion from a slab source as shown in Eq. S6.

$$I(x) = a\left[\text{erf}\left(\frac{x+h}{\sqrt{4D_b t + 4\sigma^2}}\right) - \text{erf}\left(\frac{(x-h)}{\sqrt{4D_b t + 4\sigma^2}}\right)\right] + I_b \qquad (S6)$$

I(x) is the normalized intensity of dopants at depth x, a is the proportional constant, h is the thickness of the film, σ is the SIMS mixing parameter, $I_b$ is the normalized background intensity, and $D_b$ is the bulk diffusion coefficient. The fitted results for grain boundary and bulk diffusivities obtained from SIMS profiles for both Al and Mg are shown in **Fig.** S16. Below we provide the fitted equations for each diffusion coefficient alongside the calculated diffusivities extrapolated to 850°C for comparison. We note that the activation energies for diffusion along the grain boundaries are $\frac{1}{2}$ and $\frac{1}{5}$ of the bulk values (for $Al_2O_3$ 1.3 eV vs 2.6 eV and MgO 0.84 eV vs 4.1 eV) and the differences in diffusivities for the bulk vs grain boundaries at a common temperature, e.g. 850°C, through extrapolation is about 5 orders of magnitude ($D_{GB,Al/Mg,850} \sim 10^{-16} cm^2/s$, while $D_{Bulk,Al/Mg,850} \sim 10^{-21} cm^2/s$).

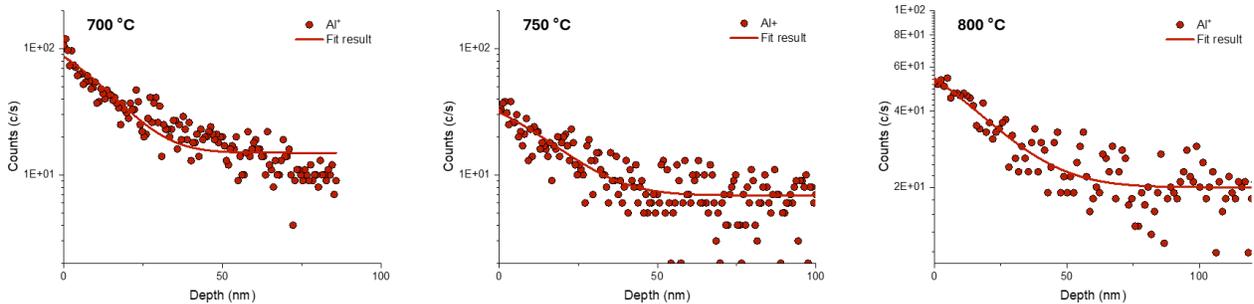

***Figure** S12: Normalized depth profiles of Al+ in GDC film layer for samples annealed at 700, 750, and 800 °C for 9, 6, and 6 hours. The best-fit lines are illustrated as red curves in each spectrum. Annealing time was adjusted to induce the desired diffusion amount and improve the profile fitting quality.*

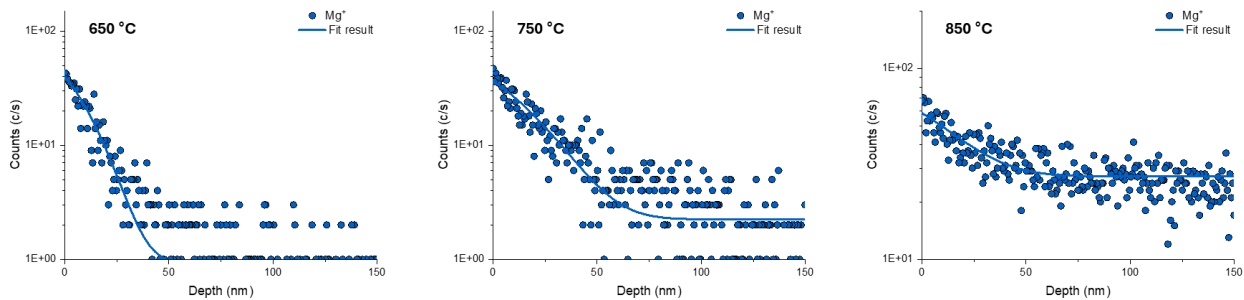

***Figure** S13: Normalized depth profiles of $Mg^+$ in GDC film layer for samples annealed at 650, 750, 850 °C for 6 hours. The best-fit lines are illustrated as blue curves in each spectrum.*

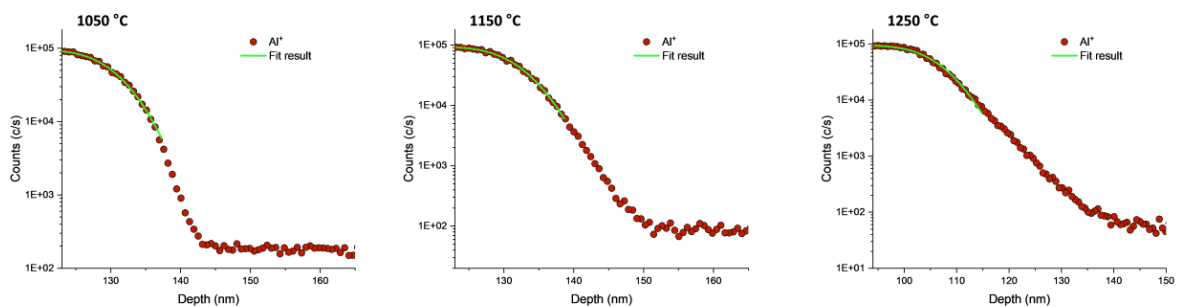

***Figure** S14: Normalized depth profiles of $Al^+$ in GDC film layer for samples annealed at 1050, 1150, 1250 °C for 30 hours. The best-fit curves are illustrated as red lines in each spectrum.*

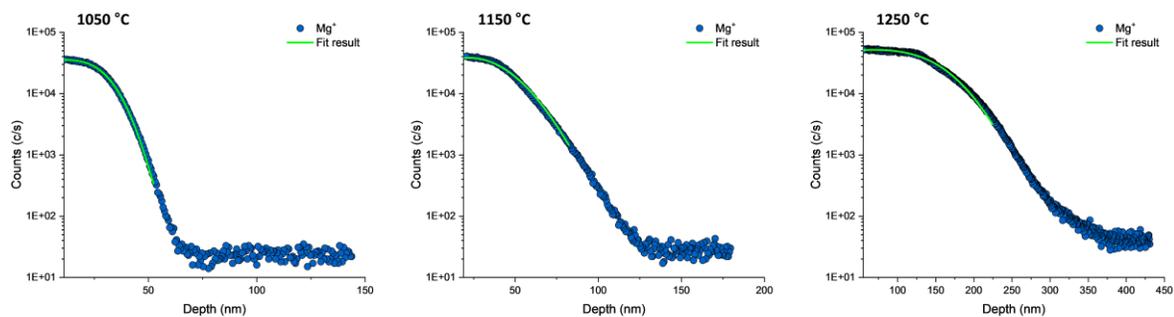

***Figure** S15: Normalized depth profiles of $Mg^+$ in GDC film layer for samples annealed at 1050, 1150, 1250 °C for 30 hours. The best-fit curves are illustrated as blue lines in each spectrum.*

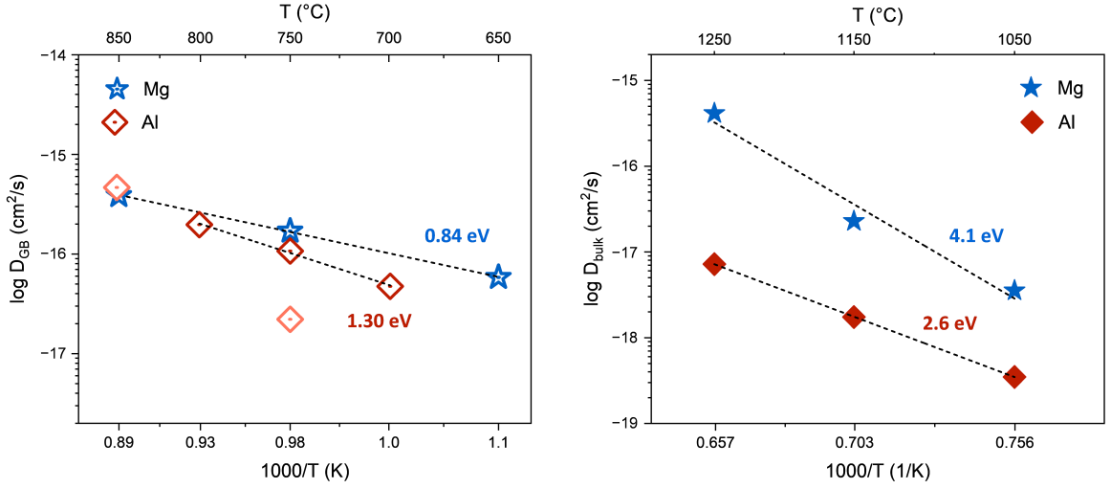

***Figure*** *S16: Arrhenius plots of measured Mg and Al diffusivities with fitted activation energies indicated. Grain boundary diffusivities on left and bulk diffusivities on right.*

**Equations for Grain Boundary Diffusivity and Bulk Diffusivity (cm$^2$.s$^{-1}$)**

$D_{GB,Al} = 6.87 \times 10^{-10} \times e^{-\frac{15057}{T}}$

$D_{GB,Mg} = 2.44 \times 10^{-12} \times e^{-\frac{9817}{T}}$

$D_{bulk,Al} = 3.63 \times 10^{-9} \times e^{-\frac{30525}{T}}$

$D_{bulk,Mg} = 1.24 \times 10^{-2} \times e^{-\frac{47641}{T}}$

**Diffusivity (cm$^2$.s$^{-1}$) at 850 °C**

$D_{GB,Al,850} = 1.11 \times 10^{-15}$

$D_{bulk,Al,850} = 6.57 \times 10^{-21}$

$D_{GB,Mg,850} = 4.08 \times 10^{-16}$

$D_{bulk,Mg,850} = 5.85 \times 10^{-21}$

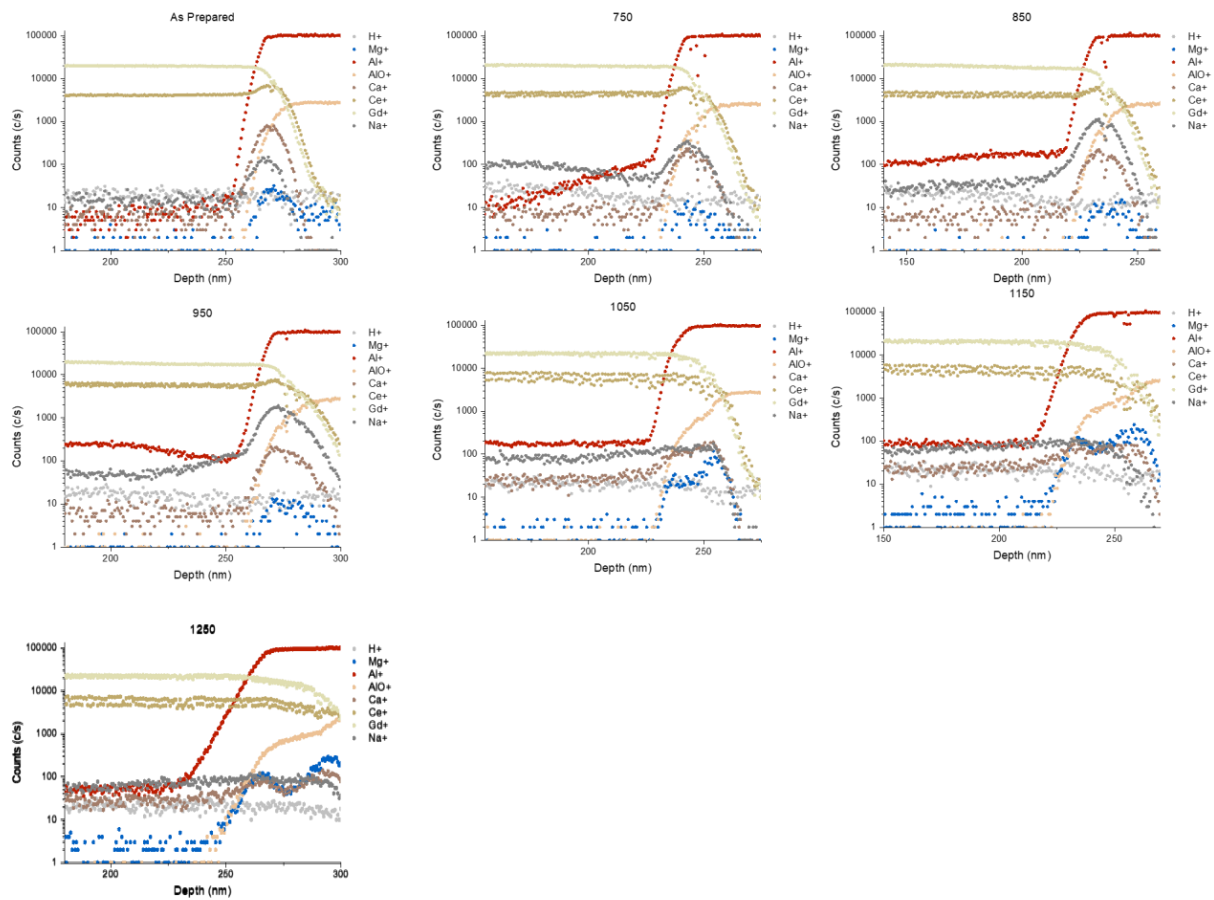

*Figure* S17: *ToF-SIMS spectra profiles of cations for GDC deposited on $Al_2O_3$. Depth profiles were obtained for as-deposited samples and samples annealed at varied temperatures for 6 hours.*

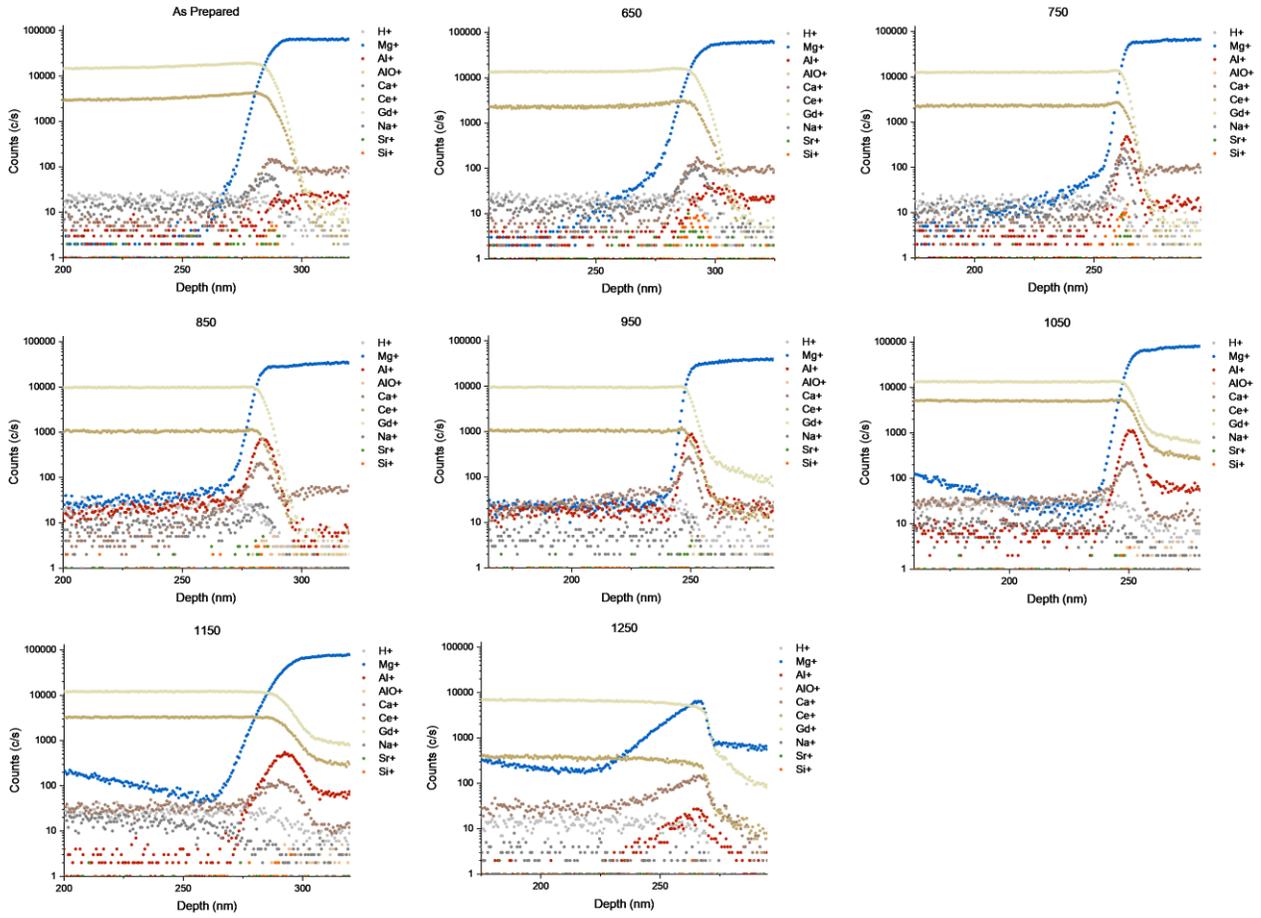

*Figure S18: ToF-SIMS spectra profiles of cations for GDC deposited on MgO. Depth profiles were obtained for as-deposited samples and samples annealed at varied temperatures for 6 hours.*

## 9. STEM-EDS Dataset Collected from Grain Boundaries of GDC3 Grown on Al$_2$O$_3$ and MgO

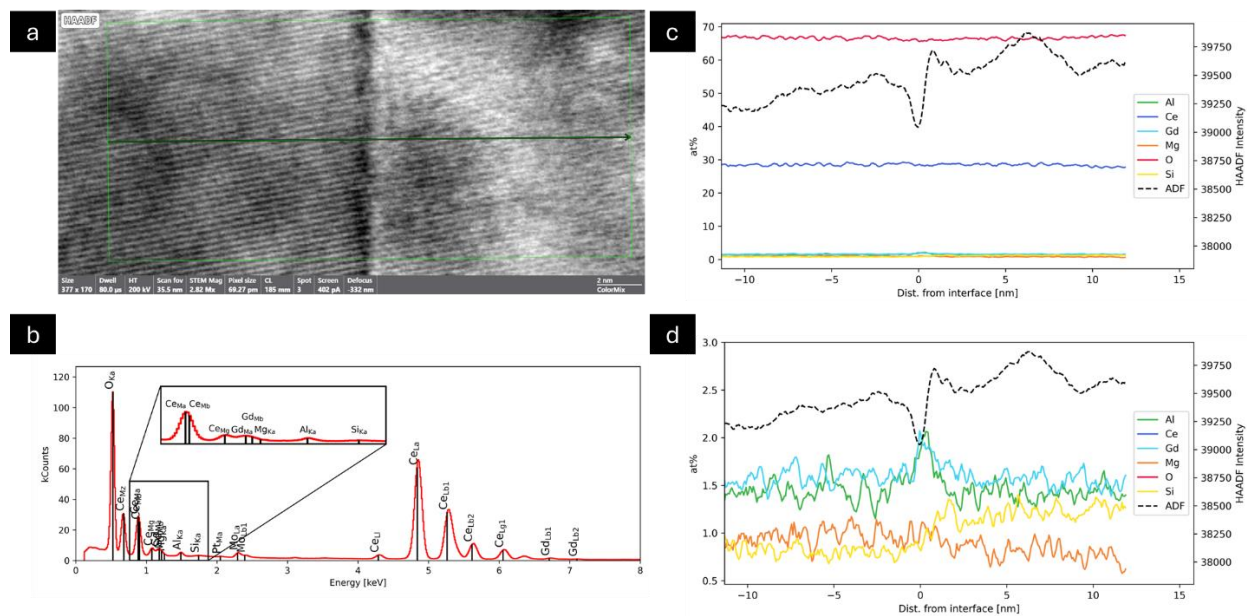

*Figure* S19: *STEM-EDS dataset collected from a representative grain boundary of GDC grown on Al$_2$O$_3$ and annealed at 850°C. a) High-angle annular dark field (HAADF-STEM) overview of dataset. The grain boundary travels vertically through the center of the image. The green box and arrow indicate the linescan displayed in C-D. Scalebar 2 nm. b) Summed STEM-EDS spectrum. Al peak is visible. Mo peaks are artifacts from the sample grid, and Pt peaks result from sample preparation. c,d) Semi-quantitative concentrations measured across the grain boundary in a) and HAADF intensity with d) showing zoomed-in intensities to display trace elements. Al and Gd are enriched at the grain boundary.*

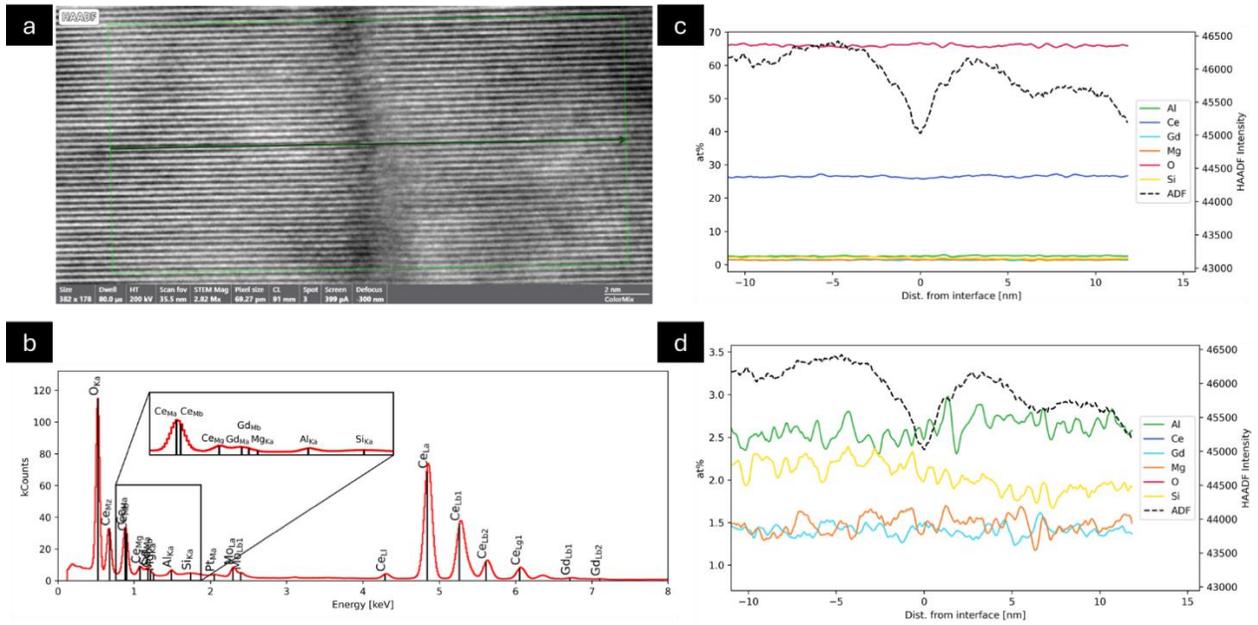

*Figure S20: STEM-EDS dataset collected from a grain boundary of GDC grown on $Al_2O_3$ and annealed at 850°C, which displays no enrichment. a) High-angle annular dark field (HAADF-STEM) overview of dataset. The grain boundary travels vertically through the center of the image. The green box and arrow indicate the linescan displayed in C-D. Scalebar 2 nm. b) Summed STEM-EDS spectrum. Al peak is visible. Mo peaks are artifacts from the sample grid, and Pt peaks result from sample preparation. c,d) Semi-quantitative concentrations measured across the grain boundary in a) and HAADF intensity with d) zoomed to display trace elements. No evidence of chemical changes are found at the grain boundary.*

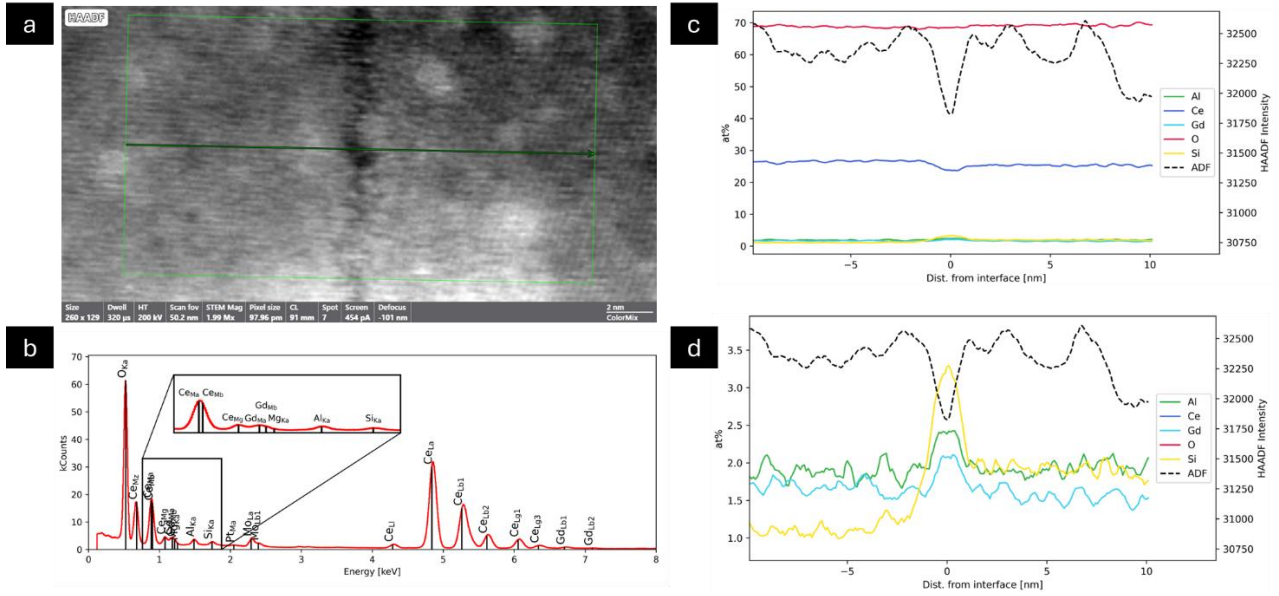

*Figure S21:* STEM-EDS dataset collected from a representative grain boundary of GDC grown on $Al_2O_3$ and annealed at 1050°C. High angle annular dark field (HAADF-STEM) overview of dataset. Grain boundary travels vertically through the center of the image. Green box and arrow indicate the linescan displayed in C-D. Scalebar 2 nm. b) Summed STEM-EDS spectrum. Mg, Al, and Si peaks are visible. Mo peaks are artifacts from the sample grid, and Pt peaks are a result of sample preparation., c,d) Semi-quantitative concentrations measured across the grain boundary in a), as well as HAADF intensity. d) is zoomed to display trace elements. Al and Si are enriched at the grain boundary, and there is weak enrichment of Gd.

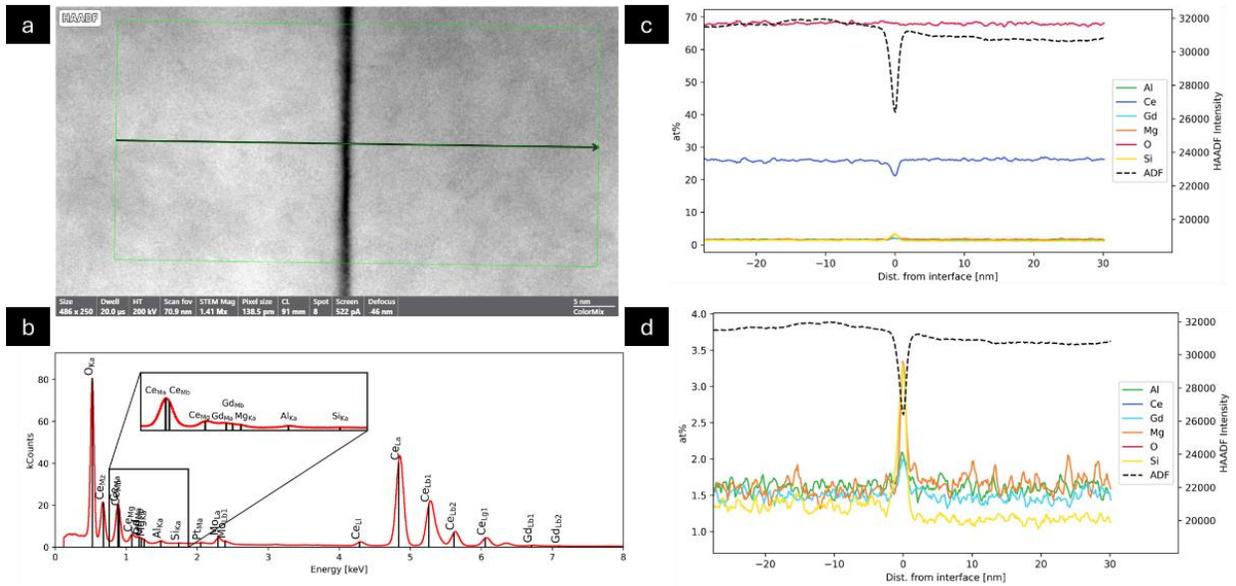

*Figure S22:* STEM-EDS dataset collected from a representative grain boundary of GDC grown on MgO and annealed at 1050°C. a) High angle annular dark field (HAADF-STEM) overview of dataset. Grain boundary travels vertically through the center of the image. Green box and arrow indicate the linescan displayed in c-d. Scalebar 5 nm. b) Summed STEM-EDS spectrum. Mg, Al, and Si peaks are visible. Mo peaks are artifacts from the sample grid, and Pt peaks are a result of sample preparation., c,d) Semi-quantitative concentrations measured across the grain boundary in a), as well as HAADF intensity with d) zoomed-in to display trace elements. Mg and Si are strongly enriched at the grain boundary. Weak enrichment of Al and Gd is also evidenced.

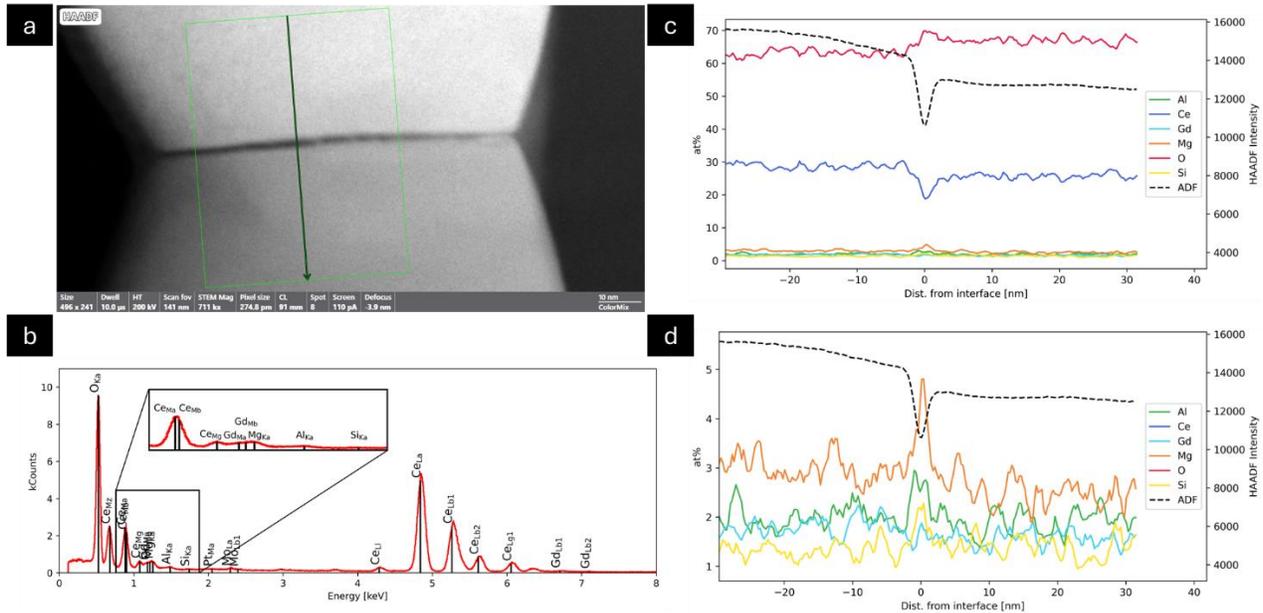

*Figure S23:* STEM-EDS dataset collected from a representative grain boundary of GDC grown on MgO and annealed at 1250°C. a) High angle annular dark field (HAADF-STEM) overview of dataset. Grain boundary travels horizontally through the image. MgO substrate is visible on the right of image. Green box and arrow indicate the linescan displayed in C-D. Scalebar 10 nm. b) Summed STEM-EDS spectrum. Mg and trace Al are visible. Mo peaks are artifacts from the sample grid. c,d) Semi-quantitative concentrations measured across the grain boundary in a), as well as HAADF intensity. There is some evidence of oxygen enrichment at the grain boundary. d) is zoomed-in to display trace elements. Mg is clearly enriched at the grain boundary. No evidence of Gd enrichment is found.

*Note on Si segregation at grain boundaries:* From Fig. S21-S23 we observe Si segregation to the grain boundaries, but only at elevated temperatures (>1000°C). Its impact on GB resistance remains debatable. Some literature suggests interstitial incorporation due to its small size (radius 0.4 Å – 6 fold coordination), imparting a positive core charge and increasing the boundary resistance due to an increase in space charge potential[5], while others indicate the possible formation of insulating siliceous phases, also increasing boundary resistance, but gettering impurities leading to decreased potential barriers[11]. As shown in Fig. 2 of main manuscript, we do not observe increases in resistance at higher temperatures, while we do observe decreases in space charge potential. This would be indicative of the second scenario (i.e. gettering interface impurities). At the same time, given that the resistance did not increase, this suggests that any "siliceous phases" formed would be present only in non-continuous patchy form. Furthermore, clear differences in resistance between the series of thin films on $Al_2O_3$ vs those on MgO remain, even at higher temperatures. This indicates that if Si is playing a gettering role, it is only partial and does not significantly impact the observed conductance trends which we assign to the presence of Mg and Al.

## 10. High Temperature Restricted Equilibrium Scenario

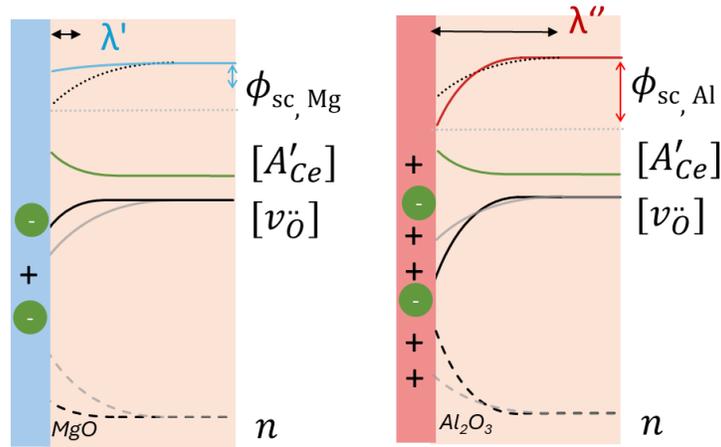

*Figure S24: Physical model explaining spatial distributions of space charge potential, bulk dopant concentration, majority oxygen vacancy, and minority electron defect concentrations in response to combined substrate cation up-diffusion and bulk dopant Gd segregation. As samples are annealed at higher temperatures, the grain boundaries become saturated with Al (right) and Mg (left) respectively, while ultimately, Gd segregation at the grain boundary further reduces the space charge potential for samples on MgO, while counteracting the increase observed with GB decoration by Al in samples on Al₂O₃. The concentration profile and space charge width follow a restricted equilibrium scenario instead of simple mott Schottky approximation.*

**Fig**. S24 describes the expected impact of competition between substrate element up-diffusion, at lower temperatures, and bulk dopant segregation to the grain boundary core occurring at higher temperatures. This situations leads to a modified restricted equilibrium scenario where the space charge width becomes smaller, proportional to the Debye length ( $L_D = \left(\frac{kT\varepsilon}{2e^2[Gd'_{Ce,bulk}]}\right)^{0.5}$ ) and less sensitive to the space charge potential itself.